\documentclass{aa}  

\usepackage{amsmath}
\usepackage{tablefootnote}
\usepackage{bm}
\usepackage{xcolor}
\usepackage[normalem]{ulem}
\usepackage{graphicx}
\usepackage{placeins}

\usepackage{txfonts}
\usepackage[colorlinks,linkcolor=blue,citecolor=blue,linktocpage=true,breaklinks, plainpages=false,urlcolor=blue]{hyperref}

\usepackage{rotating}
\usepackage{sidecap}
\usepackage{subcaption}
\usepackage{url}
\usepackage{soul}
\hypersetup{allcolors=blue}
\usepackage{float}
\usepackage[tracking=true,kerning=true,shrink=50]{microtype}
\usepackage{lipsum}
\usepackage{txfonts}
\usepackage[export]{adjustbox} 
\bibpunct{(}{)}{;}{a}{}{,}
\DeclareUnicodeCharacter{00A0}{~}
\usepackage[us,12hr]{datetime}
\usepackage{ulem}

\begin{document} 

    \title{Investigating the Period-Luminosity Relations of $\delta$ Scuti Stars: A Pathway to Distance and 3-D Dust Map Inference}

    \author{Fangzhou Guo
        \inst{1}
        \and
        Joshua S. Bloom
        \inst{2,3}
        \and
        Xiaofeng Wang
        \inst{1}
        \and
        Liyang Chen
        \inst{1}
        \and
        Jie Lin
        \inst{4}
        \and
        Xiaodian Chen
        \inst{5}
        \and
        Jun Mo
        \inst{1}
        \and
        Jicheng Zhang
        \inst{6}
        \and
        Shengyu Yan
        \inst{1}
        \and
        Qichun Liu
        \inst{1}
        \and
        Haowei Peng
        \inst{1}
        \and
        Xiaojun Jiang
        \inst{5}
        \and
        Xiaoran Ma
        \inst{1}
        \and
        Danfeng Xiang
        \inst{7,1}
        \and
        Wenxiong Li
        \inst{8}
        }
    
    \institute{
    Physics Department, Tsinghua University, Beĳing, 100084, People's Republic of China
    \and
    Astronomy Department, University of California, Berkeley, CA 94720, USA 
    \and
    Lawrence Berkeley National Laboratory, 1 Cyclotron Road, MS 50B-4206, Berkeley, CA 94720, USA
    \and
    Department of Astronomy, University of Science and Technology of China, Hefei, 230026, China
    \and
    CAS Key Laboratory of Optical Astronomy, National Astronomical Observatories, Chinese Academy of Sciences, Beijing 100101, China
    \and
    Department of Physics and Astronomy, Beijing Normal University, Beijing 100875, China
    \and
   Beĳing Planetarium, Beĳing Academy of Sciences and Technology, Beijing, 100044, China
   \and
   National Astronomical Observatories, Chinese Academy of Sciences, Beijing 100101, China
    \\
    \email{joshbloom@berkeley.edu; wang\_xf@mail.tsinghua.edu.cn}
    }

   \date{Draft: compiled on \today\ at \currenttime~UT}
   \authorrunning{Guo et al.}
   \titlerunning{Investigating the $P$-$L$ Relations of $\delta$ Scuti Stars}

    \abstract
    {
    While $\delta$ Scuti stars---intermediate-mass stars pulsating with periods $<0.3$ d---are the most numerous class of $\kappa$-mechanism pulsators in the instability strip, the short periods and small peak-to-peak amplitudes have left them understudied and underutilized. Recently, large-scale time-domain surveys have significantly increased the number of identified $\delta$ Scuti stars, allowing for more comprehensive investigations into their properties. Notably, the Tsinghua University–Ma Huateng Telescopes for Survey (TMTS), with its high-cadence observations at 1-minute intervals, has identified thousands of $\delta$ Scuti stars, greatly expanding the sample of these short-period pulsating variables. 
    } 
    {
    This study makes use of multiband photometric time-series data to refine the period-luminosity ($P$-$L$) relations of $\delta$ Scuti stars and show how observed $P$-$L$ relations can be used to simultaneously infer dust obscuration and distance. Using spectroscopy, we also study the dependence of the $P$-$L$ relations on metallicity.
    }
    {
    Using the $\delta$ Scuti stars from the TMTS catalogs of Periodic Variable Stars, we cross-matched the dataset with Pan-STARRS1, 2MASS, and WISE to obtain photometric measurements across optical ($g$, $r$, $i$, $z$, and $y$), near-infrared ($J$, $H$, $K_s$) and mid-infrared ($W1$, $W2$, $W3$) bands, respectively. Parallax data, used as Bayesian priors, were retrieved from $Gaia$ DR3, and line-of-sight dust extinction priors were estimated from a three-dimensional dust map. Using \texttt{PyMC}, we performed a simultaneous determination of the 11-band $P$-$L$ relations of $\delta$ Scuti stars. 
    }
    {
    The simultaneous determination of multiband $P$-$L$ relations of $\delta$ Scuti stars not only yields precise measurements of these relations, but also greatly improves constraints on the distance moduli and color excesses, as evidenced by the reduced uncertainties in the posterior distributions. Furthermore, our methodology enables an independent estimation of the color excess through the $P$-$L$ relations, offering a potential complement to existing 3-D dust maps. Moreover, by cross-matching with LAMOST DR7, we investigated the influence of metallicity on the $P$-$L$ relations. Our analysis reveals that incorporating metallicity might reduce the intrinsic scatter at longer wavelengths. However, this result does not achieve 3$\sigma$ significance, leaving open the possibility that the observed reduction is attributable to statistical fluctuations.
    }
    {We introduced an innovative approach to studying the $P$-$L$ relations of $\delta$ Scuti stars, facilitating more comprehensive investigations into their utility as distance indicators and their significance in understanding stellar evolution. Our extensible methodology also enables the inference of dust extinction using pulsating stars beyond $\delta$ Scuti stars. Although the inclusion of metallicity in the $P$-$L$ relations appears to reduce intrinsic scatter at longer wavelengths, further analysis is required to fully understand the impact of metal abundance on properties of $\delta$ Scuti stars. 
    }
    
    
    \maketitle

\nolinenumbers
\modulolinenumbers[0]

% -----------------------------------------------------------------------
\section{Introduction}
\label{sec:intro}
In the Hertzsprung–Russell (H-R) diagram, $\delta$ Scuti stars exist at the intersection of the classical instability strip and the zero-age main sequence (ZAMS, \citealt{breger1979delta,rodriguez2001delta}). Characterized by short periods (< 0.3 d) and small peak-to-peak amplitudes, $\delta$ Scuti stars are $A$0 to $F$5-type intermediate-mass stars \citep{chang2013statistical}. They represent a transitional class, with temperatures ranging from 6,900 to 8,900 K as well as masses between 1.5 $M_{\odot}$ and 2.5 $M_{\odot}$. These pulsators bridge the gap between low-mass stars that exhibit radiative cores and convective envelopes, and high-mass stars that are characterized by prominent convective cores and radiative envelopes \citep{bowman2017amplitude}, making them crucial for studying stellar structure and evolution history. In addition, $\delta$ Scuti stars occupy a transitional role in stellar pulsation, linking large-amplitude radial pulsators like Cepheids with non-radial, multi-periodic oscillators \citep{breger2000delta}.

Like other well-studied pulsating variables, namely Cepheids and RR\,Lyrae stars, the pulsation of $\delta$ Scuti stars is mainly excited by the $\kappa$ mechanism \citep{baker1965pulsations,breger1979delta}, where the opacity varies as a function of temperature and density in the \hbox{He II} partial ionization zones near 48,000 K \citep{chevalier1971short}. Most $\delta$ Scuti stars are multimode pulsators, with both radial ($l$=0) and non-radial ($l$=1,2,3,...) modes, as well as fundamental ($n$=1) and overtone ($n$=2,3,4,...) modes \citep{uytterhoeven2011kepler}, which has been discussed in many studies \citep{breger2000delta,guzik2019properties}. $\delta$ Scuti stars with amplitude above 0.3 mag in the $V$ band are classified as high-amplitude $\delta$ Scuti stars (HADS, \citealt{balona2016combination}), while others are typical $\delta$ Scuti stars (DSCT). HADS and DSCT are different in many aspects, including pulsation mode, rotational speed, and others \citep{pigulski2006high}. 

Adhering to the Leavitt Law (also known as the period-luminosity [$P$-$L$] relations), pulsating stars serve as crucial tools in the determination of cosmic distances. Among these, the $P$-$L$ relations of Cepheids have been the most extensively studied and utilized for decades, forming the cornerstone of our understanding of extragalactic distance scales \citep{leavitt1912periods,madore1991cepheid}. These relations have also paved the way for groundbreaking discoveries in astrophysics, including exploration of the accelerated expansion of the universe through observations of Cepheids and Type Ia supernovae \citep{riess1998observational,riess2004type}, and the constraint on the Hubble constant \citep{pierce1994hubble,riess2018milky}. As members of Population II stars, RR\,Lyrae stars also exhibit well-defined $P$-$L$ relations, making them particularly valuable for tracing the distance to old populations, such as globular clusters \citep{cacciari2003globular,braga2015distance} and dwarf galaxies in the Local Group \citep{vivas2022variable,nagarajan2022rr}.

\begin{figure}
    \centering
    \includegraphics[width=0.5\textwidth]{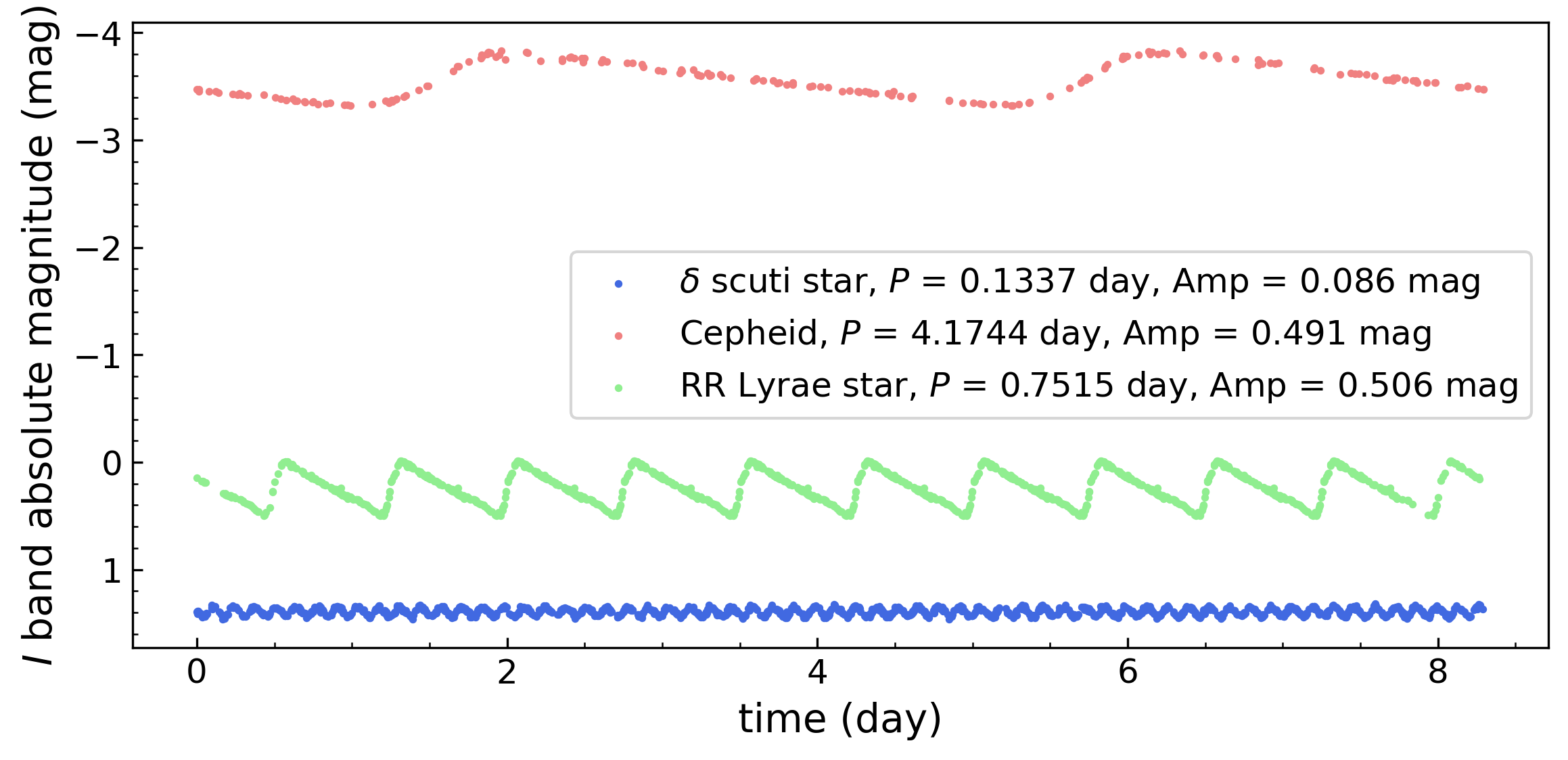}
    \caption{Comparison of the light curves of $\delta$ Scuti stars \citep{soszynski2021over}, Cepheids \citep{udalski2018ogle} and RR Lyrae stars \citep{soszynski2014over} obtained from the OGLE Collection of Variable Stars.}
    \label{fig:compare_3}
\end{figure}

Although $\delta$ Scuti stars also follow the $P$-$L$ relations, they are relatively less utilized as distance indicators due to their intrinsically fainter luminosities, which limit their applicability for extragalactic studies, and their small amplitudes ranging from 0.003 to 0.9 mag \citep{breger1979delta} that make them more challenging to detect in time-domain surveys. As shown in Figure \ref{fig:compare_3}, $\delta$ Scuti stars exhibit lower absolute magnitudes and smaller amplitudes compared to Cepheids and RR Lyrae stars. Their pulsation modes are also more complex, resulting in less tight $P$-$L$ relations. Despite these challenges, $\delta$ Scuti stars represent the second most populous class of pulsating stars in the Galaxy, surpassed only by pulsating white dwarfs \citep{breger1979delta}. This abundance, combined with their broad distribution across the Galactic disk and bulge, suggests a potential for a significant role in studying Galactic structure, stellar evolution, interstellar dust, and the calibration of distances to globular clusters \citep{mcnamara2011delta}. 

Over the past decade, the number of known $\delta$ Scuti stars has grown substantially thanks to the rapid advancement of modern large-scale time-domain surveys, and their $P$-$L$ relations have been also increasingly studied in recent years \citep{mcnamara1997luminosities,barac2022revisiting,martinez2022segmented}. \cite{chen2020zwicky} discovered \textasciitilde 15,000 $\delta$ Scuti stars in Zwicky Transient Facility (ZTF) Catalog of Periodic Variable Stars, while the All-Sky Automated Survey for SuperNovae (ASAS-SN) Catalogue of Variable Stars contains \textasciitilde 8,400 $\delta$ Scuti stars \citep{jayasinghe2020asas}. Transiting Exoplanet Survey Satellite (TESS) found \textasciitilde 15,000 $\delta$ Scuti stars with its 30-minute cadence survey \citep{gootkin2024new}. The largest collections of $\delta$ Scuti stars so far have been provided by the Optical Gravitational Lensing Experiment (OGLE), which identified over 24,000 $\delta$ Scuti stars in the Galactic bulge and disk \citep{soszynski2021over}, as well as 15,000 in the Large Magellanic Cloud \citep{soszynski2023ogle}.  

The growing availability of high-precision parallax measurements from \textit{Gaia} has revolutionized this field, allowing for unprecedented accuracy in the calibration of the $P$-$L$ relations, thus enhancing their utility in various astrophysical applications. Looking ahead, the Legacy Survey of Space and Time (LSST) is expected to overcome some of the current observational challenges associated with $\delta$ Scuti stars. With its unparalleled photometric precision (\textasciitilde 10 mmag) and sky coverage (18,000 deg$\mathrm{^2}$, \citealt{abell2009lsst}), LSST will facilitate the detection of faint and low-amplitude sources such as $\delta$ Scuti stars, potentially unlocking broader applications for these stars in the future. 

In this paper, we study the $P$-$L$ relations of $\delta$ Scuti stars in the TMTS catalogs of $\delta$ Scuti stars (part of the TMTS catalogs of Periodic Variable Stars, \citealt{guo2024minute}). The paper is structured as follows. The multiband observation data are shown in Section \ref{sec:data}, and the $P$-$L$ relations of $\delta$ Scuti stars are discussed in Section \ref{sec:plr}. In Section \ref{sec:app}, we analyze the methodology used to apply the multiband $P$-$L$ relations to estimate the color excess. Section \ref{sec:plmr} studies the influence of metallicity for the $P$-$L$ relations. We summarize our work in Section \ref{sec:summary}.

% -------------------------------------------------------
\section{Data}
\label{sec:data}
% -------------------------------------------------------

% -------------------------------------------------------
\subsection{TMTS Catalogs of $\delta$ Scuti Stars}
\label{sec:catalog}
% -------------------------------------------------------
\begin{figure}
    \centering
    \includegraphics[width=0.5\textwidth]{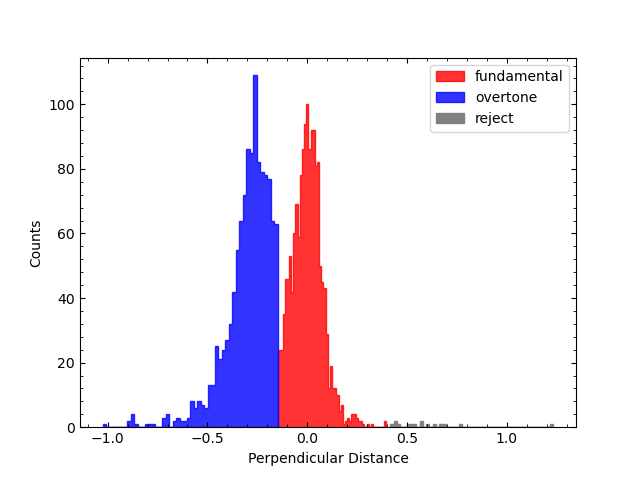}
    \caption{Distribution of the perpendicular distances to the initial fundamental mode Wesenheit $W_{JK}$ $P$-$L$ Relation for different types of pulsators. Fundamental mode pulsators are depicted in red, overtone-mode variables in blue, and candidates excluded from the analysis are shown in grey.}
    \label{fig:class}
\end{figure}

The Tsinghua University–Ma Huateng Telescopes for Survey (TMTS) is a multitube telescope system (40 cm optical telescope $\times$ 4) located at the Xinglong Station of NAOC, with a wavelength coverage of 400 to 900 nm. The TMTS has a field of view (FoV) up to 18 deg$\mathrm{^2}$ \citep{zhang2020tsinghua}. Exposing for 60s, it can reach \textasciitilde 19.4 mag in the white-light band (TMTS Luminous filter, 3$\sigma$ limitation). Having started monitoring the LAMOST sky areas since 2020, the TMTS has covered an area of about 7,000 deg$\mathrm{^2}$ during the first three years of the survey. Due to its cadence as short as 1 minute, this instrument demonstrates significant potential to identify short-period variables. Such a high cadence is particularly suitable for the detection of $\delta$ Scuti stars, especially those with periods < 2h, where rapid monitoring is essential to capture their variability. In TMTS catalogs of Periodic Variable Stars \citep{guo2024minute}, we employed machine learning techniques to classify 11,638 periodic variable stars into 6 main types, including $\delta$ Scuti stars, eclipsing binaries and candidates of RS Canum Venaticorum stars. The subset of $\delta$ Scuti stars in this catalog, known as the TMTS catalogs of $\delta$ Scuti 
stars, includes 4,876 DSCT and 628 HADS, which provide the data analyzed in this study.

To enhance the robustness of our analysis, we applied relatively strict selection criteria to our sample. Following the methods of \cite{lin2023minute} and \cite{guo2024minute}, we adopted the Lomb-Scargle Periodogram (LSP; \citealt{lomb1976least,scargle1982studies}) to analyze the light curves collected by the TMTS. We calculated the maximum powers in LSP ($\mathrm{LSP}{\mathrm{_{max}}}$) and retained only sources with $\mathrm{LSP}{\mathrm{_{max}}}$ $>$ 10$\sigma$ threshold. Furthermore, we cross-matched our samples with $Gaia$ Data Release 3 (DR3, \citealt{prusti2016gaia,vallenari2023gaia}) using a matching radius of 2.0$^{\prime \prime}$ to obtain parallax measurements($\varpi$) and included only sources with $\varpi/\sigma_\varpi \geq 10.0$. 

Since $\delta$ Scuti stars exhibit multimode pulsations, with different pulsation modes adhering to distinct $P$-$L$ relations, the following step was to carefully classify the variable stars according to their pulsation modes. We obtained the color excess of each source from the three-dimensional (3D) dust maps using the \texttt{DUSTMAPS} package (map version = "bayestar19", mode = "mean", \citealt{green2018dustmaps,green20193d}). With the help of CDS-Xmatch service, we performed a cross-match with the Two Micron All Sky Survey (2MASS, \citealt{cutri2003vizier,skrutskie2006two}) and AllWISE Data Release from the Wide-field Infrared Explorer (WISE, \citealt{wright2010wide,cutri2021vizier}) with a matching radius of 2.0$^{\prime \prime}$. 2MASS obtained full-sky photometric data in three infrared bands: $J$ (1.235 $\mu$m), $H$ (1.662 $\mu$m), and $K_s$ (2.159 $\mu$m). WISE also gathered full-sky photometry across four infrared bands, and we selected the three shorter wavelength bands: $W1$ (3.368 $\mu$m), $W2$ (4.618 $\mu$m), $W3$ (12.082 $\mu$m).

We then derived the Wesenheit indexes \citep{madore1982period,lebzelter2018new}, as the $P$-$L$ relations in the absolute Wesenheit magnitudes are better defined \citep{soszynski2005optical}:
\begin{equation}
    W_{JK} = K_{S} - 0.686(J-K_{S}).
	\label{eq:W}
\end{equation}

HADS predominantly pulsate in the fundamental mode, while DSCT displays multimode pulsations, oscillating in both the fundamental mode and various overtone modes \citep{mcnamara2011delta}. Therefore, we first established a fundamental mode $P$-$L$ relation using HADS in our catalog. Outliers in the HADS samples were examined, and sources with low signal-to-noise ratios (SNR) were excluded to enhance the reliability of the relation. 

For DSCT, we calculated the perpendicular distances of each star from this fundamental mode $P$-$L$ relation. The resulting distribution of these distances is shown in Figure \ref{fig:class}, which reveals a distinct bimodal structure, as in \cite{ziaali2019period}, \cite{jayasinghe2020asas} and \cite{barac2022revisiting}: a primary peak near 0, indicative of the fundamental mode; and a secondary peak around $-$0.26, associated with the overtone mode. Based on these findings, we classified DSCT into three categories: fundamental mode pulsators (with distances between $-$0.146 and 0.4), overtone mode pulsators (with distances less than $-$0.146), and rejected pulsators (with distances greater than 0.4). Examination of the rejected pulsators revealed that these sources typically possess low-quality light curves, which limits their reliability for further analysis. Nonetheless, accurately differentiating between various overtone pulsators (e.g., first-overtone, second-overtone, etc.) only through light curve analysis proves challenging. However, these distinctions are essential for studying the $P$-$L$ relations, as each overtone mode relates to its own $P$-$L$ relation. To improve the precision of the determination of the $P$-$L$ relations, we therefore focused exclusively on fundamental mode \hbox{$\delta$ Scuti} stars.

We acknowledge that in the double-Gaussian distribution shown in Figure \ref{fig:class}, some overtone-mode pulsators may “invade” into the peak of fundamental-mode pulsators. This overlap contributes to deviations from a tight $P$-$L$ relation.

% -------------------------------------------------------
\subsection{Optical and Infrared Photometry}
% -------------------------------------------------------
\begin{figure}
    \centering
    \includegraphics[width=0.5\textwidth]{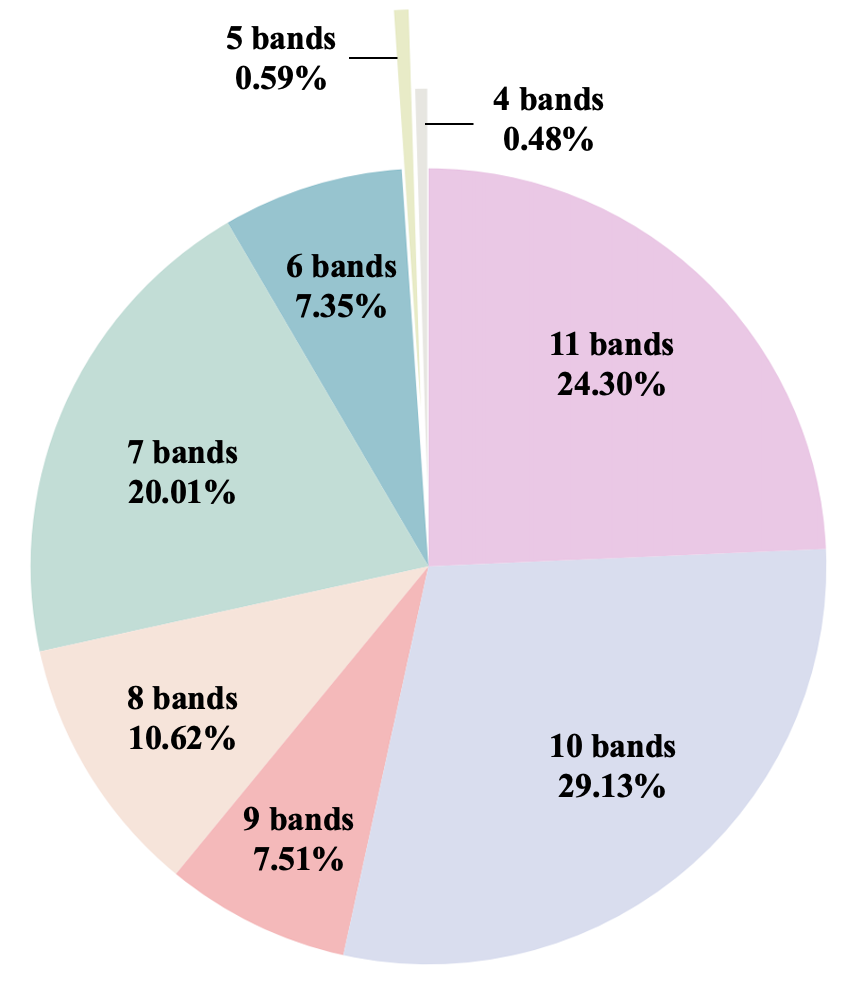}
    \caption{Number of photometric measurements of the $\delta$ Scuti stars in our dataset, as derived through the cross-matching process. More than 50\% of the sources have photometric measurements across over 10 bands.}
    \label{fig:pie}
\end{figure}

\begin{figure}
    \centering
    \includegraphics[width=0.5\textwidth]{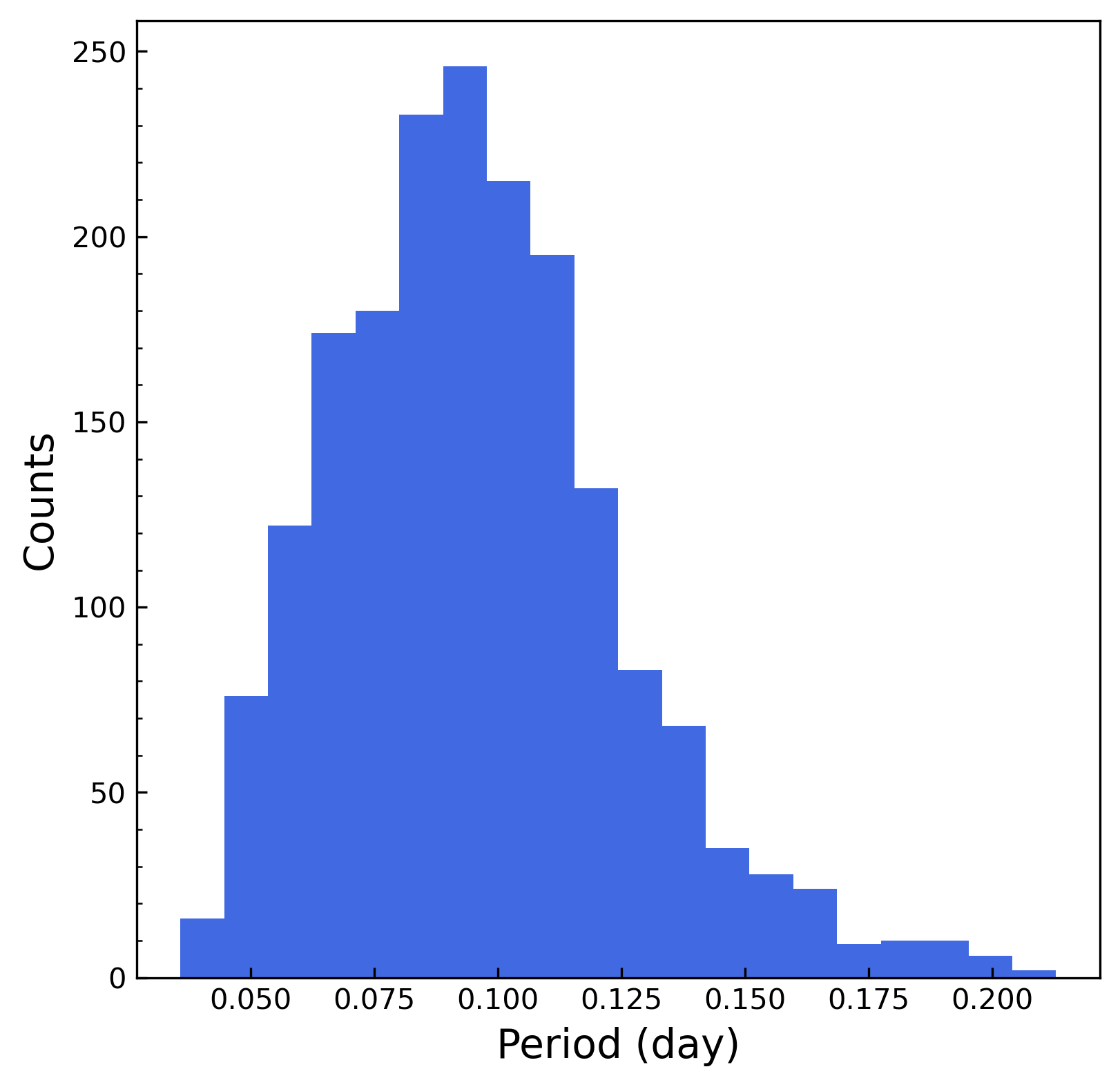}
    \caption{Periods of the $\delta$ Scuti stars in our dataset. The distribution exhibits a peak around 0.1 days (2.4 hours).}
    \label{fig:period}
\end{figure}

\begin{sidewaystable*}
	\centering
	\caption{Photometric, spectroscopic, and distance inputs to the model. The distance moduli were obtained from $Gaia$ DR3, and line-of-sight dust extinction were taken from \cite{green20193d}. Photometric measurements in the optical ($g$, $r$, $i$, $z$, and $y$), near-infrared ($J$, $H$, $K_s$) and mid-infrared ($W1$, $W2$, $W3$) bands ware derived through cross-matching with Pan-STARRS1, 2MASS and WISE, respectively. [Fe/H] values were retrieved from LAMOST DR7. See online for full electronically readable table.
}
	\label{tab:data}
	\begin{tabular}{ccccccccc} 
		\hline
            \hline
            Name & Period & $\mu_{prior}$ & $E(B-V)$ & Fe/H & $m_g$ & $m_r$ & $m_i$ & $m_z$\\
             & hour & & mag & & mag & mag & mag & mag\\
            \hline
            J00003793+5423372 & 2.2293 & 10.7010$\pm$0.0327 & 0.1622$\pm$0.0114\\					
            J00004603+5649520 & 1.7286 & 11.1031$\pm$0.0520 & 0.3250$\pm$0.0108 & 0.1360$\pm$0.0320 & 14.4009$\pm$0.0025 & 13.9847$\pm$0.0022 & 13.8352$\pm$0.0053 & 13.7514$\pm$0.0036\\				
            J00030484+5839399 & 1.7103 & 11.0794$\pm$0.0528 & 0.3650$\pm$0.0097 & & 14.2734$\pm$0.0049 & 13.8871$\pm$0.0071 & 13.7740$\pm$0.0024 & 13.6869$\pm$0.0022\\										
            J00034823+6003276 & 1.9308 & 10.8070$\pm$0.0431 & 0.3050$\pm$0.0125 & & 14.3008$\pm$0.0044 & 13.8542$\pm$0.0017 & 13.6669$\pm$0.0020 & 13.5738$\pm$0.0027\\										
            J00034856+5235145 & 3.4601 & 12.9750$\pm$0.1624 & 0.2600$\pm$0.0071 & & 14.9824$\pm$0.0027 & 14.5987$\pm$0.0014 & 14.4217$\pm$0.0036 & 14.3658$\pm$0.0031\\										
            J00044974+5859474 & 1.6179 & 10.9085$\pm$0.0548 & 0.3452$\pm$0.0233 & & 14.7557$\pm$0.0079 & 14.1761$\pm$0.0091 & 13.9080$\pm$0.0040 & 13.7521$\pm$0.0025\\										
            J00050456+5740514 & 1.6628 & 10.6668$\pm$0.0378 & 0.3400$\pm$0.0179 & & 13.8615$\pm$0.012 & 13.5173$\pm$0.0083 & 13.3328$\pm$0.0051 & \\			
            J00052274+5423014 & 2.3284 & 12.5707$\pm$0.1241 & 0.2250$\pm$0.0150 & & 14.9695$\pm$0.0021 & 14.6263$\pm$0.0042 & 14.4883$\pm$0.0025 & 14.4315$\pm$0.0015\\										
            J00052714+5457047 & 2.5555 & 12.3956$\pm$0.1394 & 0.2050$\pm$0.0139 & & 15.0639$\pm$0.0087 & 14.7264$\pm$0.0058 & 14.5995$\pm$0.0026 & 14.5610$\pm$0.0032\\	
            J00054297+5836594 & 2.3944 & 11.6862$\pm$0.0755 & 0.3681$\pm$0.0268 & & 14.8555$\pm$0.0064 & 14.3174$\pm$0.0033 & 14.1450$\pm$0.0163 & 14.0081$\pm$0.0021\\	
            J00062727+5742415 & 1.9954 & 9.9838$\pm$0.0407 & 0.2000$\pm$0.0087 & & 13.3500$\pm$0.0033 & 13.0402$\pm$0.0024\\	
            J00065315+5332565 & 2.2778 & 12.0319$\pm$0.0985 & 0.2376$\pm$0.0043 & & 14.9275$\pm$0.0029 & 14.5163$\pm$0.0045 & 14.3671$\pm$0.0035 & 14.2929$\pm$0.0113\\										
            J00080074+6344311 & 3.6114 & 12.3910$\pm$0.1006 & 0.3550$\pm$0.0287 & & 14.8023$\pm$0.008 & 14.3085$\pm$0.0057 & 14.0976$\pm$0.0055 & 14.0014$\pm$0.0074\\										
            J00081577+6239473 & 2.0532 & 11.2375$\pm$0.0768 & 0.3550$\pm$0.0130 & & 14.6689$\pm$0.0076 & 14.2743$\pm$0.0053 & 14.0259$\pm$0.0053 & 13.8763$\pm$0.0277\\										
            J00081682+5653566 & 2.2923 & 10.7963$\pm$0.0420 & 0.3600$\pm$0.0166 & & 14.1709$\pm$0.0025 & 13.6523$\pm$0.0177 & & 13.2509$\pm$0.0022\\
		\hline
	\end{tabular}\\
        \vspace{1cm}
        
        \begin{tabular}{cccccccc} 
		\hline
            \hline
            Name & $m_y$ & $m_J$ & $m_H$ & $m_{K_s}$ & $m_{W1}$ & $m_{W2}$ & $m_{W3}$\\
             & mag & mag & mag & mag & mag & mag & mag\\
            \hline
            J00003793+5423372 & 12.3932$\pm$0.0045 & 11.600$\pm$0.026 & 11.401$\pm$0.028 & 11.373$\pm$0.025 & 11.333$\pm$0.023 & 11.340$\pm$0.021 & 11.377$\pm$0.115\\
            
            J00004603+5649520 & 13.6695$\pm$0.0054 & 12.820$\pm$0.021 & 12.622$\pm$0.030 & 12.581$\pm$0.026 & 12.457$\pm$0.023 & 12.475$\pm$0.024 & 12.470$\pm$0.399\\	
            
            J00030484+5839399 & 13.5738$\pm$0.0032 & 12.744$\pm$0.024 & 12.530$\pm$0.029 & 12.413$\pm$0.029 & 12.378$\pm$0.023 & 12.383$\pm$0.025 & 12.357$\pm$0.315\\	
            
            J00034823+6003276 & 13.4670$\pm$0.0024 & 12.618$\pm$0.025 & 12.338$\pm$0.031 & 12.232$\pm$0.025 & 12.170$\pm$0.023 & 12.199$\pm$0.024 & 12.781$\pm$0.374\\	
            
            J00034856+5235145 & 14.2902$\pm$0.0026 & 13.489$\pm$0.025 & 13.209$\pm$0.026 & 13.089$\pm$0.032 & 13.046$\pm$0.023 & 13.050$\pm$0.026 & 12.938$\pm$0.520\\	
            
            J00044974+5859474 & 13.6076$\pm$0.0064 & 12.631$\pm$0.024 & 12.355$\pm$0.029 & 12.242$\pm$0.026 & 12.118$\pm$0.023 & 12.137$\pm$0.024 & 12.037$\pm$0.190\\	
            
            J00050456+5740514 & 13.2541$\pm$0.0022 & 12.404$\pm$0.027 & 12.225$\pm$0.032 & 12.113$\pm$0.025 & 12.010$\pm$0.022 & 12.028$\pm$0.022\\			
            
            J00052274+5423014 & 14.3715$\pm$0.0068 & 13.550$\pm$0.029 & 13.279$\pm$0.036 & 13.268$\pm$0.035 & 13.211$\pm$0.025 & 13.237$\pm$0.027 & 12.224$\pm$0.261\\										
            J00052714+5457047 & 14.4856$\pm$0.0040 & 13.643$\pm$0.028 & 13.428$\pm$0.039 & 13.372$\pm$0.035 & 13.306$\pm$0.024 & 13.361$\pm$0.028\\	
            
            J00054297+5836594 & 13.9126$\pm$0.0039 & 12.967$\pm$0.027 & 12.724$\pm$0.033 & 12.589$\pm$0.026 & 12.545$\pm$0.023 & 12.552$\pm$0.025\\
            
            J00062727+5742415 & 12.7278$\pm$0.0010 & 11.935$\pm$0.027 & 11.664$\pm$0.033 & 11.587$\pm$0.023 & 11.537$\pm$0.022 & 11.552$\pm$0.022 & 11.398$\pm$0.125\\	
            
            J00065315+5332565 & 14.2064$\pm$0.0024 & 13.340$\pm$0.030 & 13.148$\pm$0.032 & 12.994$\pm$0.031 & 12.965$\pm$0.025 & 12.947$\pm$0.028 & 12.485$\pm$0.326\\										
            J00080074+6344311 & 13.8759$\pm$0.0028 & 12.994$\pm$0.024 & 12.766$\pm$0.031 & 12.656$\pm$0.032 & 12.556$\pm$0.023 & 12.591$\pm$0.023 & 12.666$\pm$0.319\\										
            J00081577+6239473 & 13.8437$\pm$0.0029 & 12.824$\pm$0.022 & 12.593$\pm$0.030 & 12.489$\pm$0.029 & 12.336$\pm$0.024 & 12.340$\pm$0.024 & 12.825$\pm$0.465\\										
            J00081682+5653566 & 13.1449$\pm$0.0040 & 12.218$\pm$0.020 & 11.934$\pm$0.019 & 11.802$\pm$0.022 & 11.756$\pm$0.023 & 11.757$\pm$0.022 & 11.672$\pm$0.147\\
		\hline
	\end{tabular}    
\end{sidewaystable*}

We cross-matched the $\delta$ Scuti stars in our catalog with the Panoramic Survey Telescope and Rapid Response System (Pan-STARRS) Data Release 1 (PS1 DR1, \citealt{kaiser2002pan,chambers2016pan}) using a matching radius of 2.0$^{\prime \prime}$ for optical photometry. Pan-STARRS collected photometry in five optical bands: $g$ (4866 \AA), $r$ (6215 \AA), $i$ (7545 \AA), $z$ (8679 \AA), and $y$ (9633 \AA). The entire cross-matching process provided each source with up to 11-band photometry, including $grizy$ bands from PS1 DR1, $JHK$ bands from 2MASS and $W1-W3$ bands from ALLWISE. Table \ref{tab:data} shows a sample of our dataset. For Pan-STARRS, 2MASS and ALLWISE, each source has photometric measurements from approximately 13, 45 and 2 epochs, respectively, providing a robust temporal sampling of the light curves. The one benefit with small peak-to-peak amplitude pulsators like $\delta$ Scuti stars is that the "mean magnitude"---which can be a challenge to determine with sparse sampling of large amplitude variables---is closely measured in a single epoch. Of course, the periods are still determined from the high-precision photometry from TMTS, which continuously monitors the LAMOST sky areas throughout the night with a 1-minute cadence. This observation strategy ensures collections of uninterrupted light curves, each obtained from a single night of observations. The short cadence and uninterrupted light curves from the TMTS enable precise period determination, providing valuable constraints for the $P$-$L$ relation. 

To ensure the precision of our fitting, we selected only sources with photometry available in more than four bands. The final dataset comprises 1,864 fundamental mode $\delta$ Scuti stars, of which 1,567 are newly discovered by the TMTS (i.e., not recorded by the International Variable Star Index (VSX, \citealt{watson2006international}). A total of 16,781 photometric data were recorded. As illustrated in Figure \ref{fig:pie}, approximately 99\% of these variables have photometric measurements in six or more bands, with more than half of them having photometry in $\geq$ 10 bands. 

Figure \ref{fig:period} shows the period distribution of the sources in our dataset. Compared with other catalogs, our dataset has a larger portion of variables with extremely short periods (i.e., $P \leq$ 2 h), which adds a longer lever arm to aid in the inference of the zero points of the $P$-$L$ relations.  

\section{Period-Luminosity Relations}
\label{sec:plr}

In many studies like \cite{mcnamara2011delta} and \cite{jayasinghe2020asas} of the $P$-$L$ relations of $\delta$ Scuti stars, the $P$-$L$ relation for each band are determined separately, with the distance modulus and color excess (which are critical in calculating the absolute magnitudes) for each variable assumed to be fixed, even though they are essentially variables with probability distributions. For a $\delta$ Scuti star with a $Gaia$ parallax ($\varpi$) of 0.7 mas (the average in our dataset) and a parallax-over-error ratio of 10---corresponding to a parallax uncertainty $\sigma_\varpi$ = 0.07 mas---the resulting uncertainty in the distance modulus is $\sim$ 0.22 mag. This exceeds typical photometric uncertainties and is larger than the intrinsic scatter in most bands of the $P$-$L$ relations. Consequently, uncertainties in the distance modulus have a non-negligible impact on the determination of $P$-$L$ relations and are necessary to be accounted for in the analysis.

Using the $Python$ Package \texttt{PyMC} \citep{patil2010pymc,abril2023pymc}, we can simultaneously fit the multiband $P$-$L$ relations, extinction, and distances of the $\delta$ Scuti stars in a Bayesian context. Designed for probabilistic programming, \texttt{PyMC} has extensive support for prior distributions, MCMC sampling methods, and comprehensive model diagnostics. Within this framework, the distance modulus and the color excess due to dust are modeled as random variables with prior distributions. These priors are derived from $Gaia$ DR3 and \texttt{DUSTMAPS} \citep{green20193d}, respectively. By incorporating multiband photometry to impose cross constraints, this approach not only tightens the $P$-$L$ relations but also refines the uncertainties in the distance modulus and color excess through the posterior distributions. 

Absolute magnitude is defined as:
\begin{equation}
    M_{i,j}=m_{i,j} - \mu_i - A_j
	\label{eq:magnitude}
\end{equation}
where i refers to the $i$th source and j refers to the $j$th band. $M_{i,j}$ ($m_{i,j}$) is the absolute (apparent) magnitude of the $i$th source in the $j$th band. $\mu_i$ represents the distance modulus of the $i$th source, and $A_j$ is the interstellar extinction in the $j$th band. The Cardelli, Clayton \& Mathis (CCM) extinction law \citep{cardelli1989relationship} gives the relationship between the interstellar extinction in the $j$th band and that in the V band: 
\begin{equation}
    \langle A_j/A_V \rangle=a_j+b_j/R_V
	\label{eq:Aj}
\end{equation}
where the coefficients $a_j$ and $b_j$ are wavelength functions. Given $R_V=A_V/E(B-V)$, $A_j$ can be written as a function of the color excess $E(B-V)$:
\begin{equation}
\begin{split}
    A_j=A_V(a_j+b_j/R_V) =R_VE(B-V)(a_j+b_j/R_V)\\
       =E(B-V)(a_jR_V+b_j)
	\label{eq:ccm}
\end{split}
\end{equation}

The relationship between period and absolute magnitude (the $P$-$L$ relation) is:
\begin{equation}
    M_{i,j}=M_{0,j} + \alpha_j \log_{10}(P_i/P_0)
	\label{eq:plr-1}
\end{equation}
where $\alpha_j$ and $M_{0,j}$ are the slope and intercept of the $P$-$L$ relation in the $j$th band. $P_i$ refers to the period of the $i$th source and $P_0$ is the mean period of all $\delta$ Scuti stars in our dataset, calculated to be 2.23 hours.

Substituting Equations \ref{eq:magnitude} and \ref{eq:ccm} into Equation \ref{eq:plr-1} gives:
\begin{equation}
\begin{split}
    m_{i,j}=\mu_i + M_{0,j} + \alpha_j \log_{10}(P_i/P_0) + E(B-V)_i(a_jR_V+b_j) \\
    + \epsilon_{i,j}
	\label{eq:plr-2}
\end{split}
\end{equation}
where $\epsilon_{i,j}$ is a zero-mean independently and identically distributed Gaussian random variable, with $\epsilon_{i,j}^2 = \sigma_{m,(i,j)}^2 + \sigma_{intrinsic,j}^2$. $\sigma_{m,(i,j)}$ is the observed photometric error and $\sigma_{intrinsic,j}$ is the intrinsic scatter of the $P-L$ relation in a given band. 

The inclusion of intrinsic scatter is a common approach in Bayesian statistics to account for variability beyond the measurement uncertainties. Ideally, the scatter in observation data should be fully explained by the measurement errors, $\sigma_m$. However, in many cases, the observed dispersion exceeds what measurement uncertainties alone predict. To account for this discrepancy, an intrinsic scatter term ($\sigma_{int}$), is introduced, while this term does not include the uncertainties in distance modulus, $ E(B-V)$, or photometry. Instead, it captures model imperfection (for example, the $P$-$L$ relation of $\delta$ Scuti stars is not a perfect linear relation due to complex pulsation mechanisms and stellar rotations) and unmodeled data uncertainty (including systematic errors in photometric measurements and uncertainties in median flux determination). Without invoking such an intrinsic scatter term, several issues may arise, such as underfitting---when the model fails to capture the full dispersion of the data, leading to underestimated uncertainties---and biased parameter estimates, as the fitting process may artificially minimizes the residuals, potentially resulting in overfitting.

In the least-square fitting (LSF), the post-fit residual $\sigma _\mathrm{{LSF}}$ is computed by encompassing both measurement uncertainties and intrinsic scatter, but without distinguishing between the above two sources. In contrast, the Bayesian method explicitly incorporates intrinsic scatter as a parameter which is optimized during the fitting process, allowing for a more precise decomposition of the observed scatter into measurement errors and intrinsic variability of the $P$-$L$ relation.

As in \cite{klein2014towards}, we constructed a sparse matrix $X$ to accommodate the simultaneous fitting of the 11-band $P$-$L$ relations, where matrix multiplication is used to implement Equation \ref{eq:plr-2}: 
\begin{equation}
    m = X \cdot b,
	\label{eq:matrix}
\end{equation}
where $X$ is a 16,781 $\times$ 3,750 matrix, and each row corresponds to an individual photometric measurement. The 3,750 columns represent the parameters of interest: 1,864 distance modulus, 1,864 color excess values (corresponding to 1,864 sources), along with 11 $\alpha$ and 11 $M_0$ (corresponding to 11 bands), yielding a total of 1,864 + 1,864 + 11 + 11 = 3,750. The vector $m$ contains the 16,781 photometric measurements, while the parameter vector $b$ represents the random variables, initially assigned with prior distributions for each $\delta$ Scuti star's distance modulus and color excess, as well as for the $\alpha$, $M_0$ and intrinsic scatters of the 11 bands. 

The prior distributions are as follows: 
\begin{equation}
\begin{split}
    \mu_{i,\text{Prior}} \sim \mathcal{N}(\mu_{gaia}, \sigma_{gaia}^2)\\
    E(B-V)_{i,\text{Prior}} \sim \mathcal{N}(E(B-V)_{dustmaps}, \sigma_{dustmaps}^2),\\
    \quad E(B-V)_{i,\text{Prior}} \geq 0\\
    \alpha_{j,\text{Prior}} \sim \mathcal{N}(0, 5^2)\\
    M_{0,j,\text{Prior}} \sim \mathcal{N}(0, 3^2)
	\label{eq:distribution}
\end{split}
\end{equation}

The \texttt{DUSTMAPS} library does not directly provide the standard error of the color excess for a given coordinate. To estimate this uncertainty, we employed the "sample" mode and obtained $\sigma_{dustmaps}$ by calculating the standard error from several samples. For $\alpha$ and $M_0$, we adopted a broad normal prior distribution to avoid imposing unnecessary constraints on the estimation. 

Using \texttt{PyMC}, we trained a model to obtain posterior estimates of all random variables. The \texttt{PyMC} model ran in 5,000 steps, until all the variables converged. To assess convergence, we adopt the Gelman-Rubin diagnostic ($\hat{R}$, \citealt{gelman1992inference}). Convergence is indicated when $\hat{R}$ <1.1, suggesting that the chains have stabilized. The closer $\hat{R}$ is to 1, the stronger the evidence of convergence. $\hat{R}$ is computed as follows:
\begin{equation}
    \hat{R} = \sqrt{\frac{\left(1 - \frac{1}{n}\right) W + \left(\frac{1}{n}\right) B}{W}},
	\label{eq:rhat}
\end{equation}

\begin{figure*}
    \centering
    \includegraphics[width=1\textwidth]{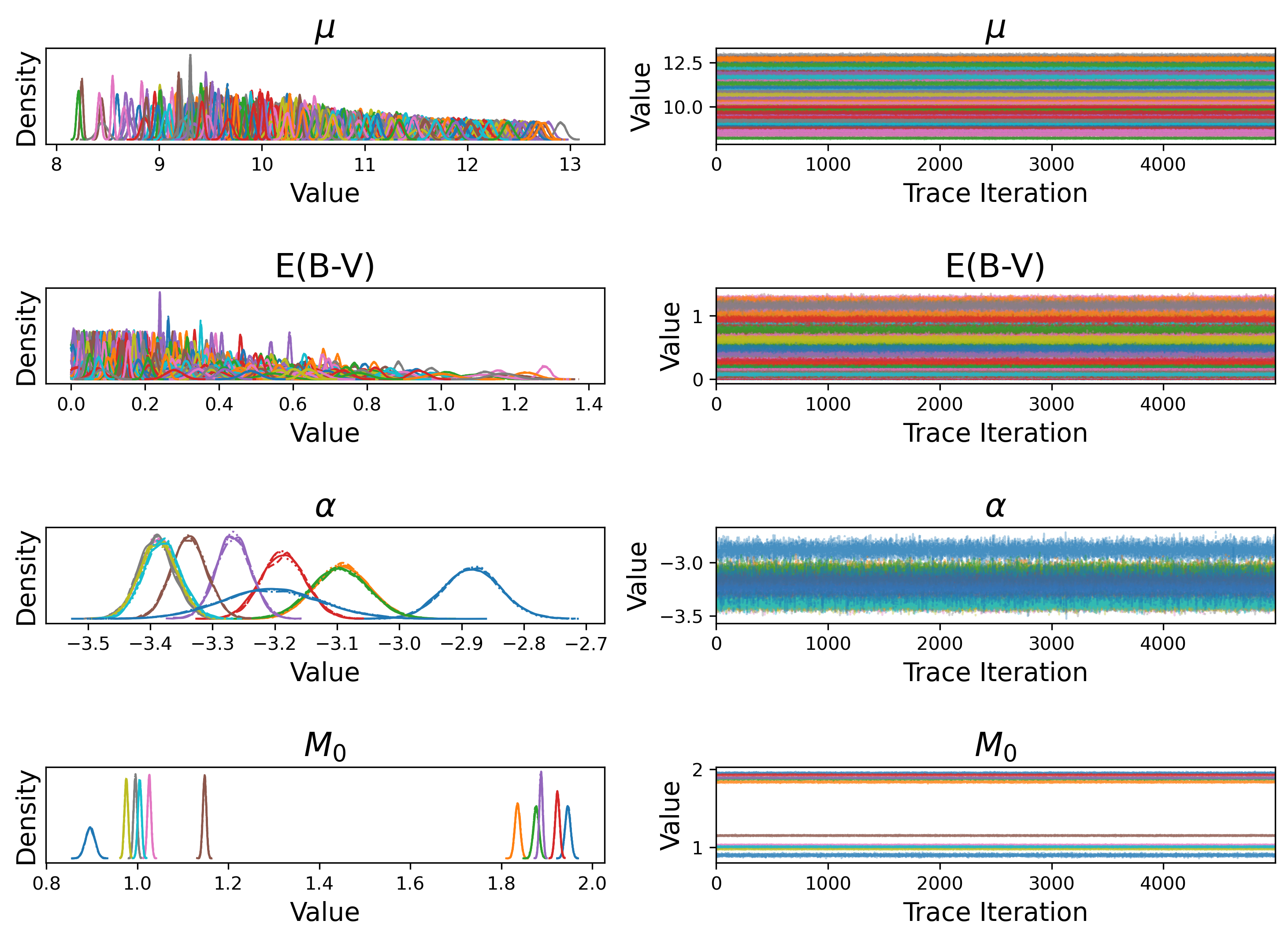}
    \caption{Posterior density plots indicating the range and relative likelihood (left) and trace plots showing the values of each parameter over the course of iterations (right) of $\mu$, $E(B-V)$, $\alpha$ and $M_0$. In the upper two panels, each color represents a distinct $\delta$ Scuti star; in the lower two panels, each color corresponds to a different photometric band. The stability of the distributions indicates that the chains have achieved satisfactory convergence.}
    \label{fig:trace_plot}
\end{figure*}

\begin{figure*}
    \centering
    \includegraphics[width=0.93\textwidth]{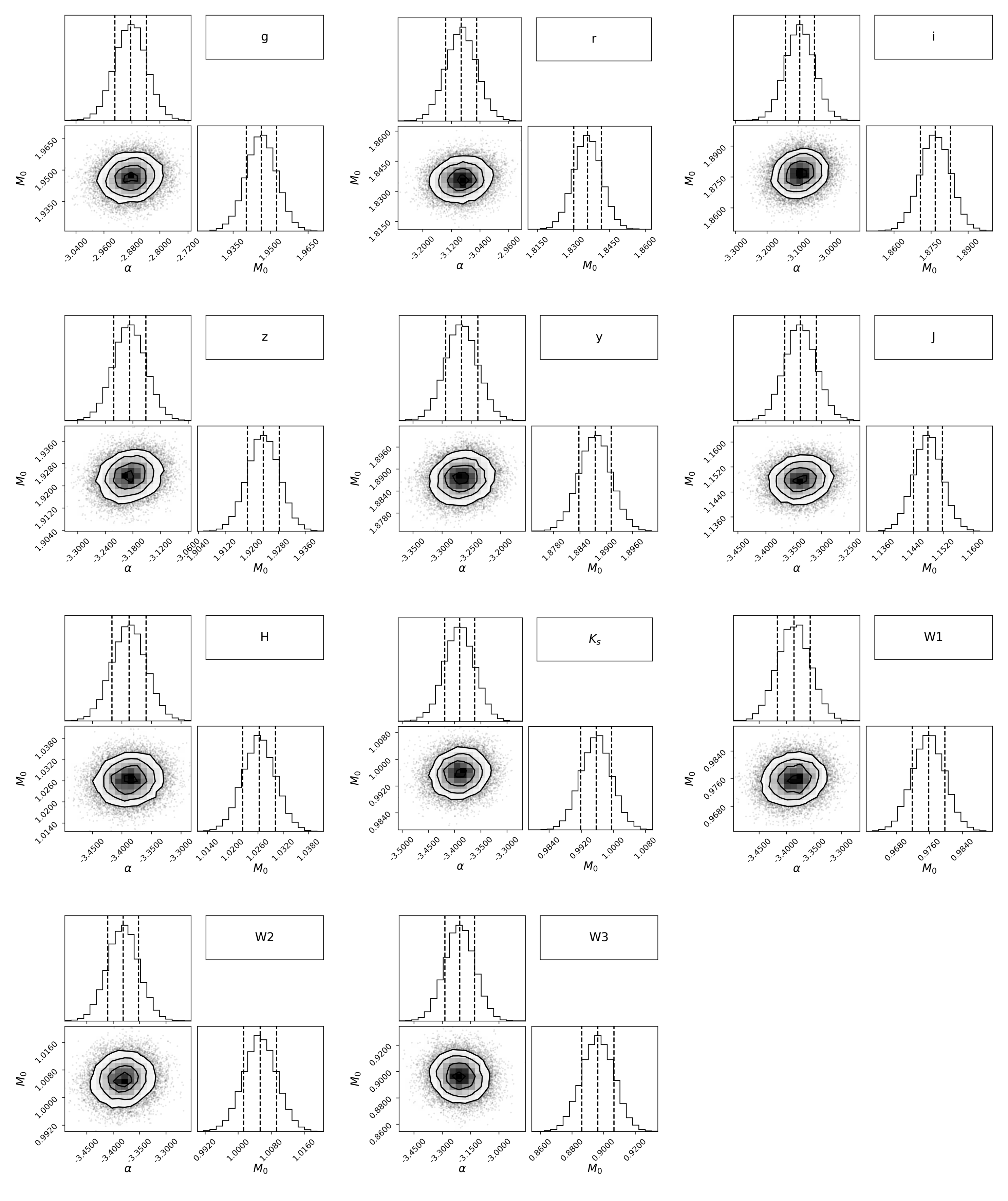}
    \caption{Contour plots for the joint posterior distributions of $\alpha$ and $M_0$ across the 11 photometric bands ($g$, $r$, $i$, $z$, $y$, $J$, $H$, $K_s$, $W1$, $W2$, and $W3$). For each band, the bottom-left panel shows the contour of $\alpha$ versus $M_0$, while the marginal posterior distributions for the two variables are displayed on the top-left and bottom-right panels, respectively. The dashed vertical lines in the marginal distributions represent the 1$\sigma$ credible intervals.}
    \label{fig:contour}
\end{figure*}

\noindent where $W$ represents the variance within the chain, $B$ is the variance between chains, and $n$ denotes the length of the chain. For our analysis, all variables yielded $\hat{R}$ = 1, indicating robust convergence. Figure \ref{fig:trace_plot} shows the trace plots and posterior density distributions for the variables in our model. For all parameters, the trace plots illustrate consistent horizontal bands without notable trends, indicating good mixing and convergence. The density plots also reveal well-defined distributions with minimal multi-modality, suggesting reliable parameter estimations. Figure \ref{fig:contour} presents the contour plots for the joint posterior distributions of $\alpha$ and $M_0$ in the 11 bands. The contour lines approach a circular shape, indicating a weak correlation between $\alpha$ and $M_0$, which implies that the model constrains these parameters relatively independently and the result is robust.

\begin{table*}
	\centering
	\caption{$\alpha$, $M_0$ and intrinsic scatters of the 11-band $P$-$L$ relations.}
	\label{tab:plr}
	\begin{tabular}{cccc} 
		\hline
		Band & $\alpha$ & $M_0$ (mag) & $\sigma_\mathrm{{intrinsic}}$ (mag)\\
		\hline
		g & $-$2.8834$\pm$0.0458 & 1.9464$\pm$0.0061 & 0.200$\pm$0.005\\
		r & $-$3.0920$\pm$0.0432 & 1.8359$\pm$0.0058 & 0.174$\pm$0.004\\
		i & $-$3.0955$\pm$0.0462 & 1.8767$\pm$0.0062 & 0.171$\pm$0.005\\
            z & $-$3.1868$\pm$0.0355 & 1.9235$\pm$0.0048 & 0.149$\pm$0.004\\
            y & $-$3.2663$\pm$0.0278 & 1.8875$\pm$0.0037 & 0.139$\pm$0.003\\
            J & $-$3.3377$\pm$0.0284 & 1.1479$\pm$0.0039 & 0.149$\pm$0.003\\
            H & $-$3.3877$\pm$0.0289 & 1.0263$\pm$0.0039 & 0.153$\pm$0.003\\
            $K_s$ & $-$3.3903$\pm$0.0290 & 0.9958$\pm$0.0039 & 0.151$\pm$0.003\\
            W1 & $-$3.3864$\pm$0.0299 & 0.9758$\pm$0.0040 & 0.157$\pm$0.003\\
            W2 & $-$3.3810$\pm$0.0298 & 1.0053$\pm$0.0041 & 0.161$\pm$0.003\\
            W3 & $-$3.2068$\pm$0.0787 & 0.8964$\pm$0.0102 & 0.298$\pm$0.002\\
		\hline
	\end{tabular}
\end{table*}

\begin{figure*}
    \centering
    \includegraphics[width=0.93\textwidth]{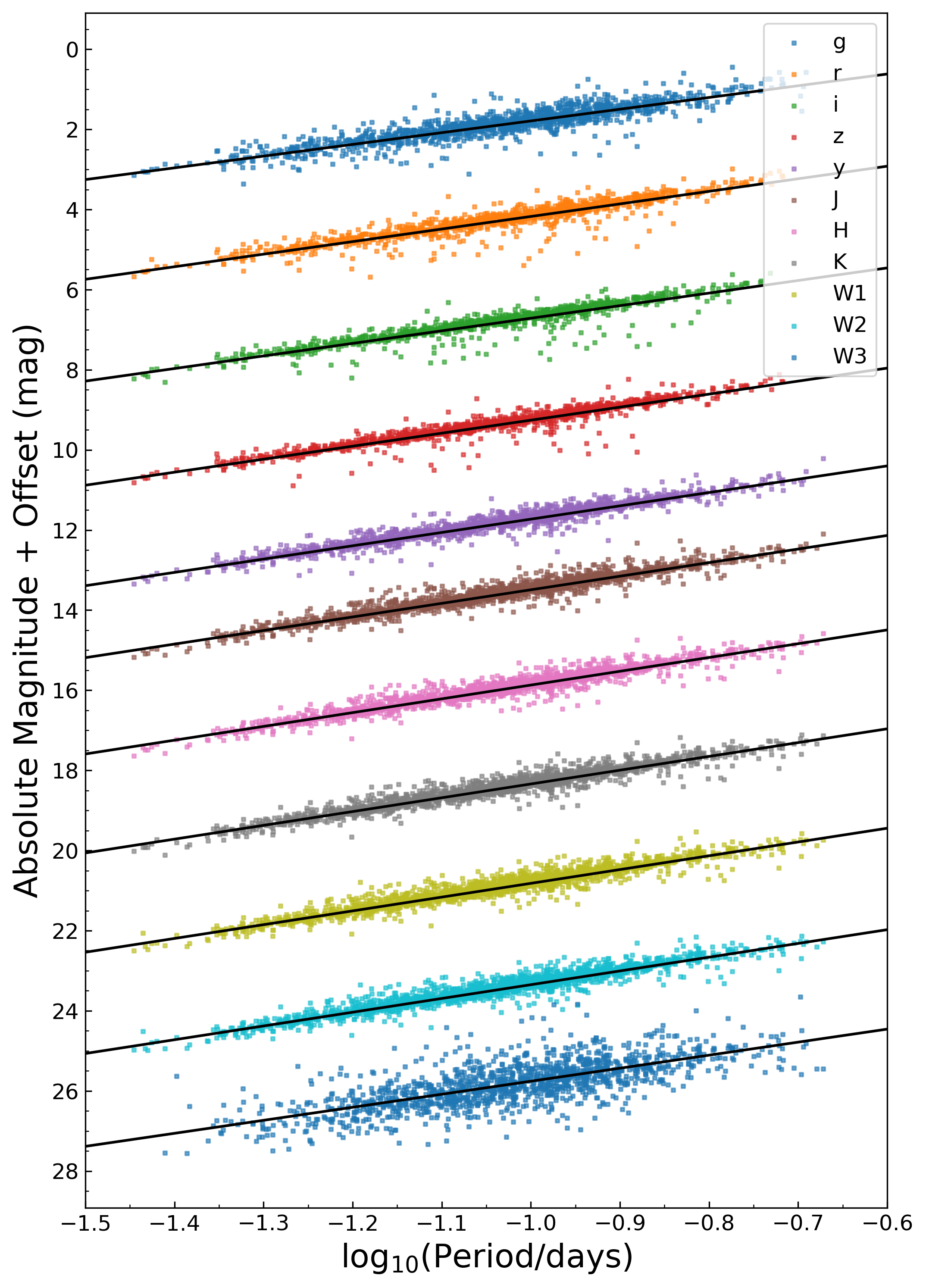}
    \caption{Period-Luminosity relations of $\delta$ Scuti stars in the 11 bands, as denoted by different colors. The zero points of the $P$-$L$ relations are vertically shifted to visually separate each relation. The solid black lines represent the best-fitting $P$-$L$ relations.}
    \label{fig:plr}
\end{figure*}

\begin{figure}
    \centering
    \includegraphics[width=0.5\textwidth]{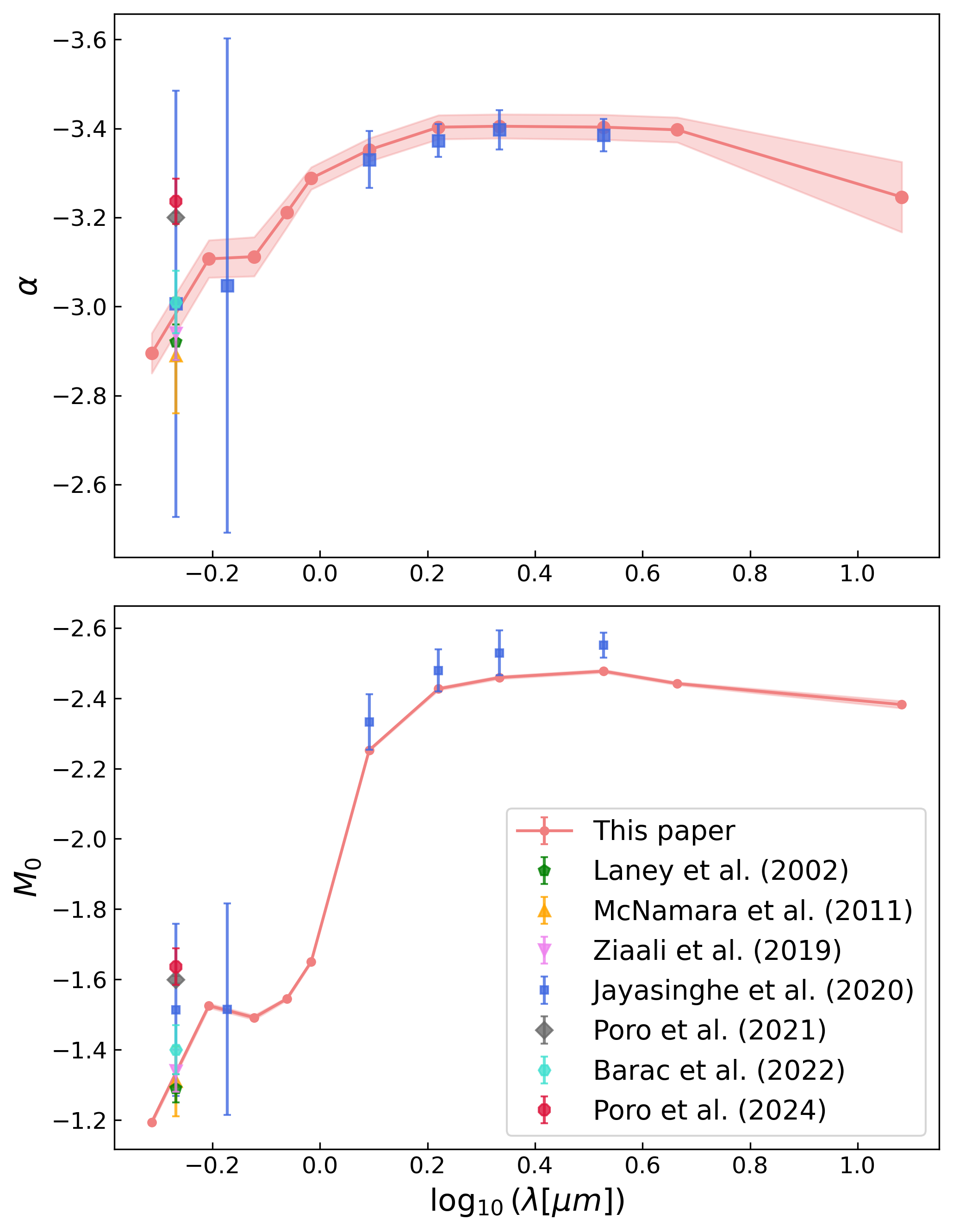}
    \caption{Comparison of the $P$-$L$ relations slope ($\alpha$) and intercept ($M_0$) of this paper with \cite{laney2002dwarf,mcnamara2011delta,ziaali2019period,jayasinghe2020asas,poro2021observational,barac2022revisiting} and \cite{poro2024period}. Our results align well with previous investigations.}
    \label{fig:alpha_comparing}
\end{figure}

\begin{figure*}
    \centering
    \includegraphics[width=1\textwidth]{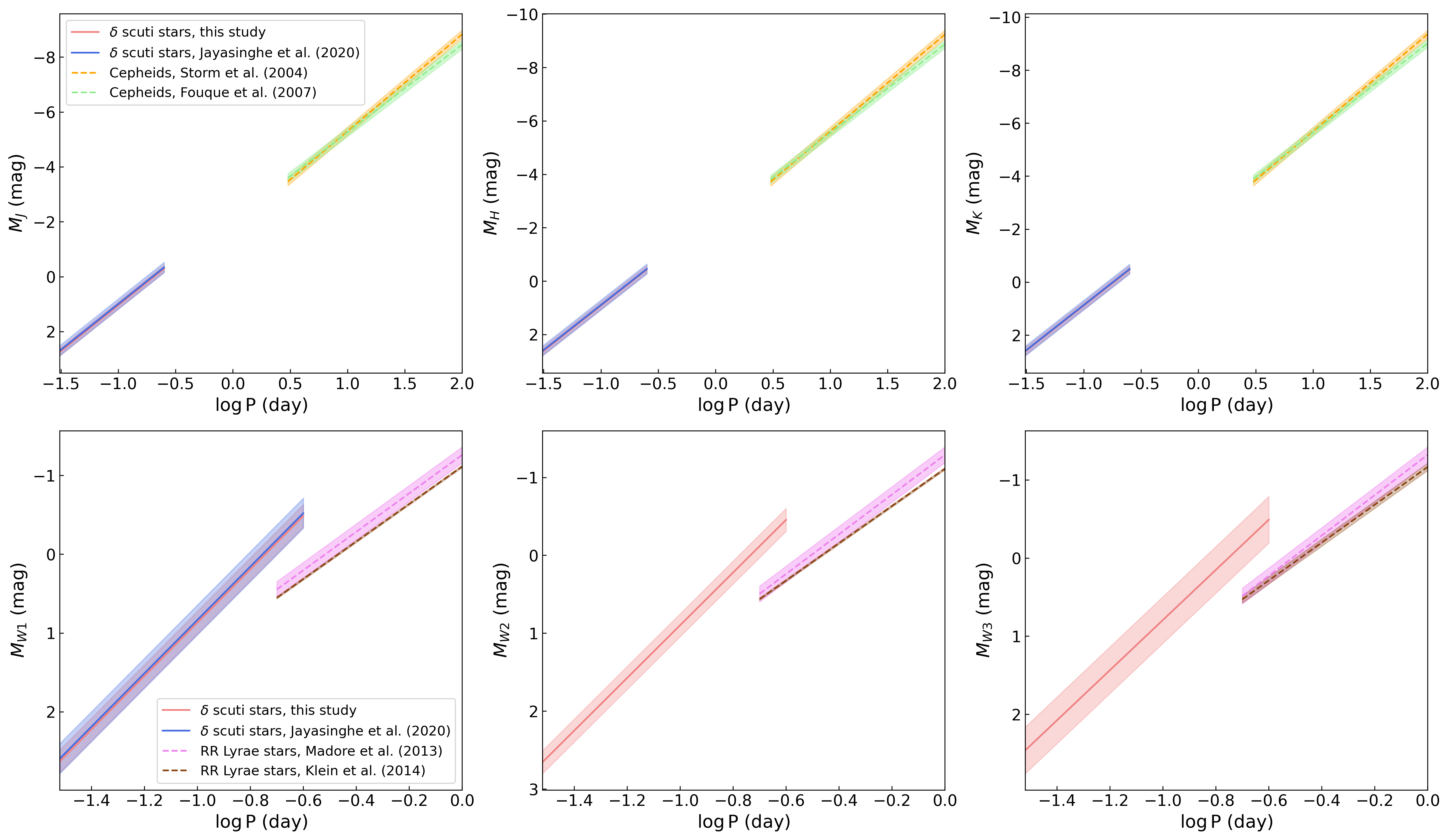}
    \caption{Comparison of the $P$-$L$ relations of $\delta$ Scuti stars and Cepheids in the $J$ (upper left), $H$ (upper middle), and $K_s$ (upper right) band, alongside the comparison of $\delta$ Scuti stars and RR Lyrae stars in the $W_1$ (lower left), $W_2$ (lower middle), and $W_3$ (lower right) band. The period range for $\delta$ Scuti stars spans 0.02–0.25 day \citep{chang2013statistical}, for Cepheids 3–100 days \citep{persson2004new}, and for RR Lyrae stars 0.2–1 day \citep{lafler1965rr}.}
    \label{fig:cepheid_comparing}
\end{figure*}

The multiband $P$-$L$ relations of $\delta$ Scuti stars are shown in Figure \ref{fig:plr}. The zero points of the $P$-$L$ relations are vertically offset to visually separate each relation, allowing for clear graphical comparison. The solid black lines represent the best-fitting $P$-$L$ relations. The fitted $\alpha$, $M_0$ and $\sigma_\mathrm{{intrinsic}}$ of each band are summarized in Table \ref{tab:plr}. Note that the scatter in the $\delta$ Scuti $P$-$L$ relation is larger than that of Cepheids and RR Lyrae stars (i.e., $\sim$0.1--0.2 mag, \citealt{fouque2007new,madore2013preliminary,dambis2014mid,trahin2021inspecting}). The pulsation modes of $\delta$ Scuti stars are complex due to multiple reasons, including multi-modal oscillation, excitation mechanism, stellar structure, and rapid rotation. All these factors contribute to the intrinsic scatter, making the $P$-$L$ relations less tight. However, it is important to note that these effects, along with the overtone "invaders" discussed in Section \ref{sec:catalog}, have been accounted for by the intrinsic scatter term introduced in our model.

Figure \ref{fig:alpha_comparing} compares our results with \cite{laney2002dwarf,mcnamara2011delta,ziaali2019period,jayasinghe2020asas,poro2021observational,barac2022revisiting} and \cite{poro2024period}. For comparison purpose, we reformulated Equation \ref{eq:plr-1} as $M_{i,j}=M_{0,j} + \alpha_j \log_{10}(P_i)$, where periods are expressed in days rather than hours. As our sample consists of Galactic $\delta$ Scuti stars, and potential differences may exist between Galactic and extragalactic $P$-$L$ relations, we have limited our comparison to studies on Galactic $\delta$ Scuti stars. Our results align well with previous studies. We further compared the $P$-$L$ relations of $\delta$ Scuti stars with Cepheids in $J$, $H$ and $K_s$ bands, and with RR Lyrae stars in $W1$, $W2$ and $W3$ bands. As shown in Figure \ref{fig:cepheid_comparing}, $\delta$ Scuti stars are characterized by intrinsically shorter periods and lower luminosities compared to Cepheids and RR Lyrae stars. These properties inherently complicate their identification and systematic study, and underscore the need for next-generation time-domain survey like LSST with better photometric precision, as discussed in Section \ref{sec:intro}.

\begin{figure*}
    \centering
    \includegraphics[width=1\textwidth]{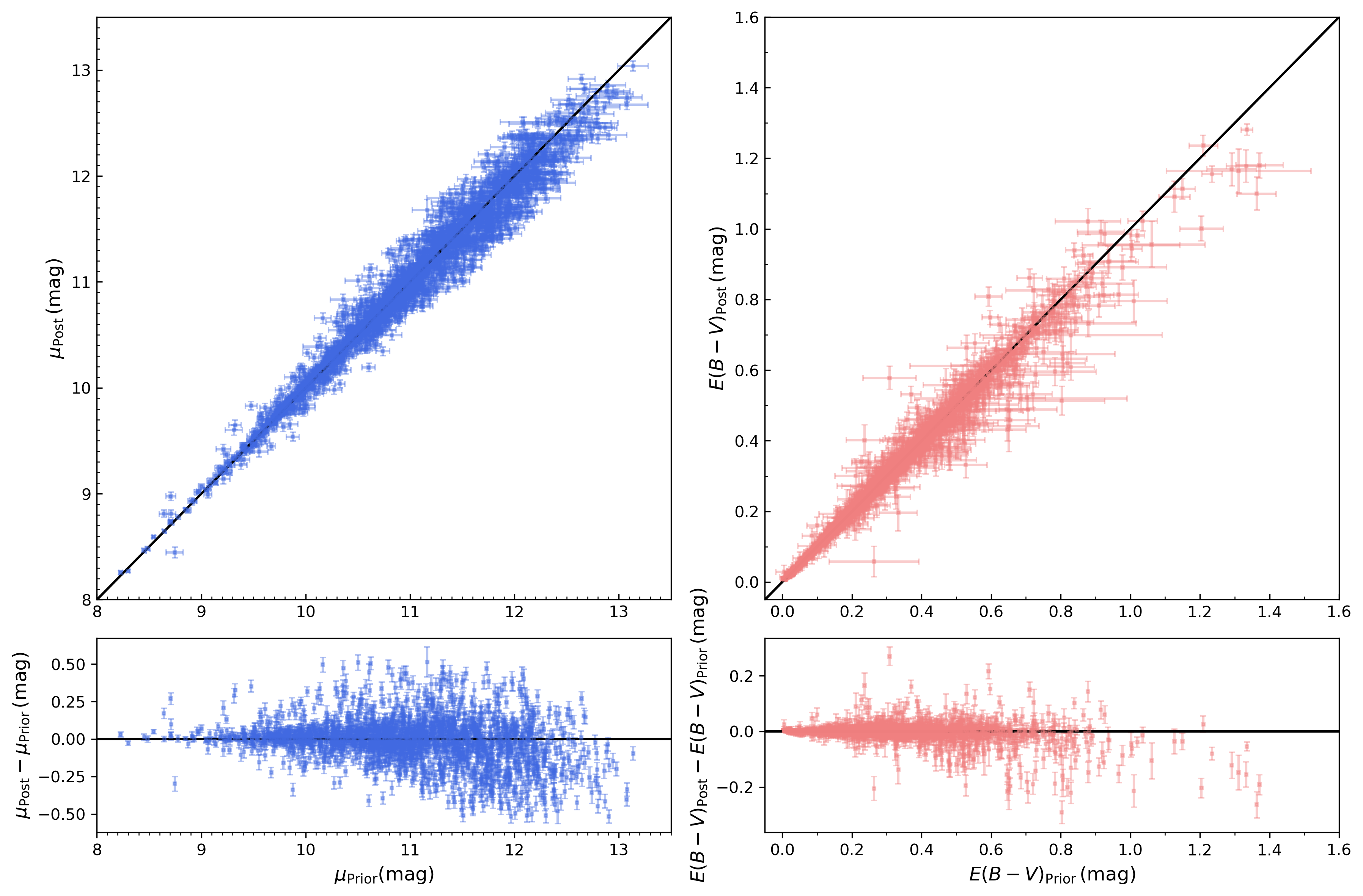}
    \caption{Comparisons of the prior and posterior distributions of distance modulus (left) and $E(B-V)$ (right), with residuals showing in the lower panels. }
    \label{fig:compare}
\end{figure*}

\begin{figure*}
    \centering
    \includegraphics[width=1\textwidth]{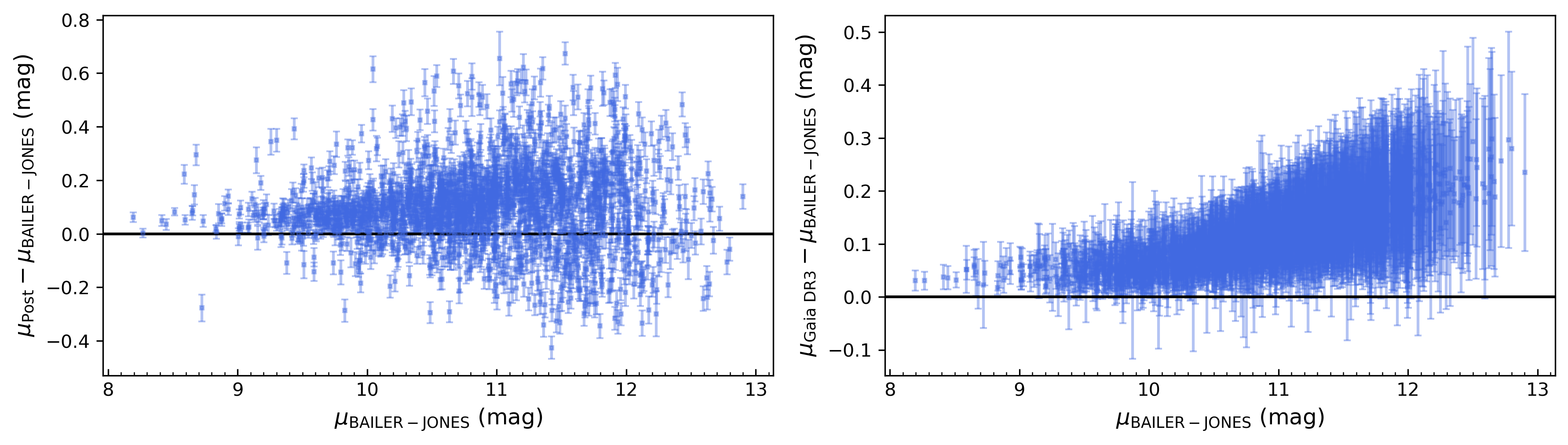}
    \caption{Comparisons of the residuals of distance modulus derived from \cite{bailer2021vizier} with our posterior distributions (left) and with $Gaia$ DR3 (right).}
    \label{fig:compare_bailer}
\end{figure*}

\begin{figure*}
    \centering
    \includegraphics[width=1\textwidth]{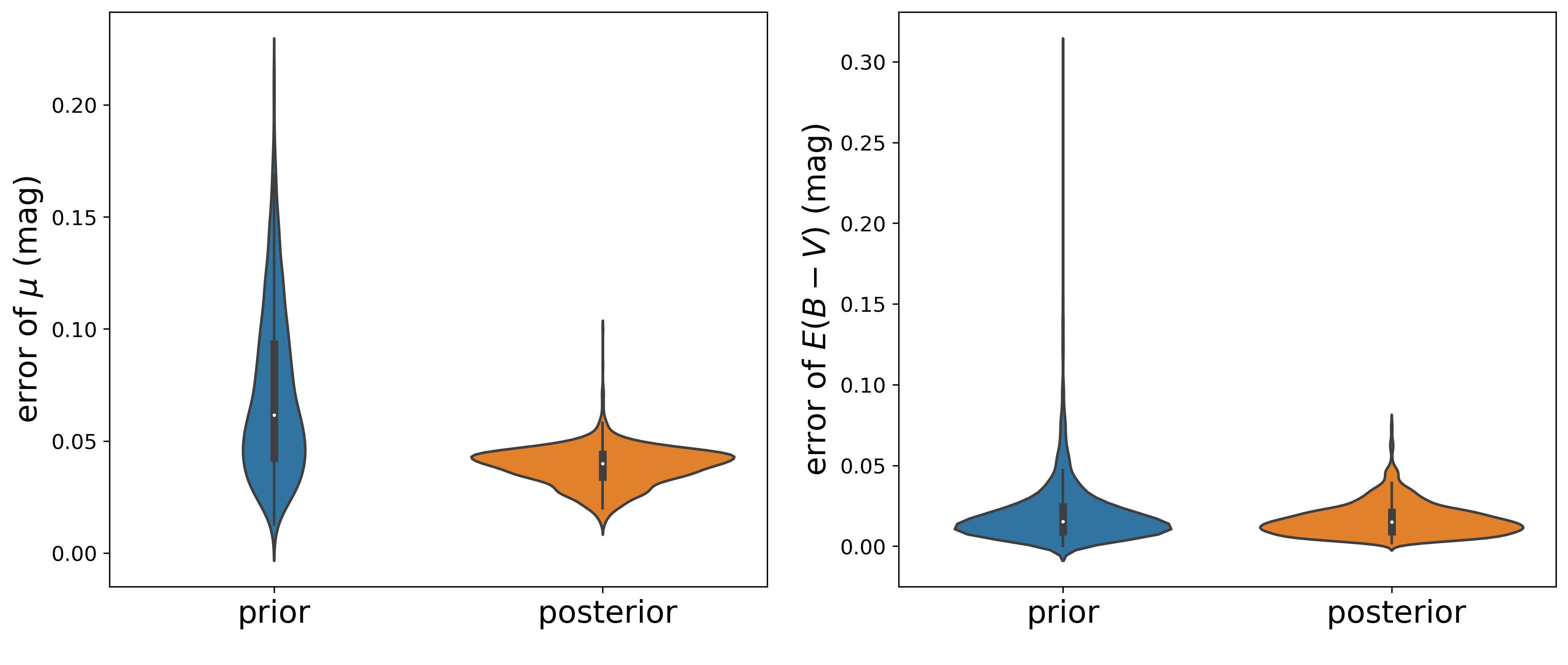}
    \caption{Violin plots depicting the prior and posterior error distribution for distance modulus (left) and $E(B-V)$ (right). The substantive depression in posterior distribution peaks relative to prior distributions shows a great error mitigation, particularly in measurements with larger initial uncertainties.}
    \label{fig:violin}
\end{figure*}

We performed a comparative analysis between the prior and posterior distributions of the distance modulus and $E(B-V)$ for the $\delta$ Scuti stars. As illustrated in Figure \ref{fig:compare}, the posterior distributions align with the priors, suggesting that the model does not introduce excessive overfitting. This alignment confirms that the model predictions are data-driven while remaining consistent with prior information. However, we observe a slight systematic deviation at larger distance, where the posterior distribution appears marginally smaller than the prior, as shown in the left bottom panel of Figure \ref{fig:compare}. To investigate this, we compared our posterior distribution with the results from \cite{bailer2021vizier}, which combines $Gaia$ DR3 parallax likelihoods with a prior based on a 3D Galactic model to improve stellar distance estimates. As shown in the left panel of Figure \ref{fig:compare_bailer}, $\mu _\mathrm{{Post}}-\mu _\mathrm{{Bailer-Jones}}$ exhibits a consistent pattern across the entire distance range, with similar trends at both the near and far ends, indicating that the observed bias discussed above disappears when the more sophisticated Galactic prior is incorporated. This is confirming evidence that the methodology does indeed lead to reasonable distance estimates. 

Note that there is another deviation in Figure \ref{fig:compare_bailer}: our posterior distances are slightly larger than those from \cite{bailer2021vizier} over the entire range. This is likely due to that the $Gaia$ DR3 distance modulus, which serve as our prior, are systematically larger than those from \cite{bailer2021vizier}, particularly at greater distances, as shown in the right panel of Figure \ref{fig:compare_bailer}. Since Bayesian methods inherently follow the prior, it is expected that our posterior---based on $Gaia$ DR3---would be systematically larger than those from \cite{bailer2021vizier}. It is worth noting that the “systematic deviation” seen in the left panel of Figure \ref{fig:compare} in turn demonstrates the effectiveness of our approach, as the distance modulus “should be” smaller under the more sophisticated Galactic prior \citep{bailer2021vizier}, especially at larger distance, and our model successfully adjusted the posterior toward this expectation.

Furthermore, we examine the reduction in uncertainties between the prior and posterior distributions of the distance modulus and $E(B-V)$, as shown in Figure \ref{fig:violin}. The comparison reveals a substantial reduction in error for both distance modulus and $E(B-V)$ in the dataset, with particularly pronounced improvements for measurements with higher uncertainties in the prior distribution. This is evident from the significantly lower peaks in the violin plots of the posterior distributions compared to the priors. Previous studies, such as \cite{sesar2017probabilistic}, have utilized the Period-Luminosity-Metallicity ($P$-$L$-$Z$) relations of pulsating stars within a Bayesian framework to constrain the parallax. This result underscores the effectiveness of our methodology in refining parameter estimates, as our framework not only preserves prior information but also incorporates observational data to provide tighter constraints on uncertainties.

% -------------------------------------------------------
\section{Application: Estimation of 3-D Dust Distribution and Distance}
\label{sec:app}
% -------------------------------------------------------

3-D dust maps provide insights into Galactic structure and enable precise corrections for dust extinction and reddening \citep{green2014measuring}. The development of 3-D dust maps has been revolutionized by advancements in large-scale, multi-band photometric surveys like 2MASS and WISE, which provide extensive photometric data on stars across various wavelengths. Combined with precise stellar distance measurements, such as those derived from $Gaia$, these surveys enable the creation of high-resolution maps that trace the spatial distribution of dust with great precision \citep{leike2020resolving,lallement2022updated}.

The implications of 3D dust maps extend beyond extinction correction. These maps deepen our understanding of Galactic structure and evolution, revealing features such as spiral arms shaped by the distribution of interstellar matter through reconstruction of the 3D distribution of dust in the Galaxy \citep{schultheis2014mapping,kh2018detection}. Dust maps also provide crucial insights into the complex interplay between interstellar dust and star formation, as grain surfaces are important catalytic sites for key chemical reactions, and dust is a fundamental component of molecular clouds where stars are born \citep{bialy2021per,dharmawardena2022three}. Furthermore, they aid in the calibration of cosmological observations by refining our models of the foreground extinction that affects extragalactic studies \citep{zasowski2019high,chiang2019extragalactic}.

Currently available 3D dust maps, such as those presented in \cite{green20193d}, \cite{lallement2022updated} and \cite{edenhofer2024parsec}, primarily rely on systematic comparisons between observed stellar colors, which are affected by interstellar extinction, and intrinsic stellar colors predicted by stellar types. By calculating color indices across multiple photometric bands, incorporating detailed stellar parameters such as effective temperature, surface gravity and metallicity, and building models through the Gaussian process — a robust framework has already been established for quantifying extinction at scale. 

A notable application of our methodology is its potential to serve as an alternative or complementary approach for estimating color excess, thereby expanding existing techniques. Our method facilitates inference of $E(B-V)$ by leveraging the multiband $P$-$L$ relations of $\delta$ Scuti stars (and can be extended to other types of pulsating stars). Using the well-characterized photometric and pulsation properties of these pulsators, our framework provides an independent avenue for deriving interstellar reddening, thereby offering reliability in the estimates of dust map.

Equation \ref{eq:plr-2} can be reformulated as:
\begin{equation}
\begin{split}
    m_{i,j} - \mu_i - M_{0,j} - \alpha_j \log_{10}(P_i/P_0) = E(B-V)_i(a_jR_V+b_j) \\
    + \epsilon_{i,j}
	\label{eq:plr-3}
\end{split}
\end{equation}

Using the fitted parameters $\alpha$ and $M_0$ for all 11 bands in combination with the photometry and distance modulus of a given source as before, we can estimate the color excess of that source using \texttt{PyMC}. To avoid overlap between the dataset used to derive the $P$-$L$ relation parameters and that used to predict the color excess, we randomly divided our dataset into a training set (70\%) and a test set (30\%). From the training set, we determined $\alpha$ and $M_0$ for each band using 5,000 sampling iterations. For posterior estimation, instead of employing a normal prior distribution as outlined in Equation \ref{eq:distribution}, we adopted a broad uniform distribution:
\begin{equation}
    E(B-V)_{i,\text{Prior}} \sim \text{Uniform}(0, 2)
	\label{eq:uniform}
\end{equation}

This approach simulates a scenario in which no prior extinction information is available, thereby reducing the influence of pre-existing assumptions. It provides a robust framework for testing the intrinsic capability of the model to independently infer $E(B-V)$, ensuring that the results are derived solely from the underlying data and the methodological principles.

\begin{figure}
    \centering
    \includegraphics[width=0.5\textwidth]{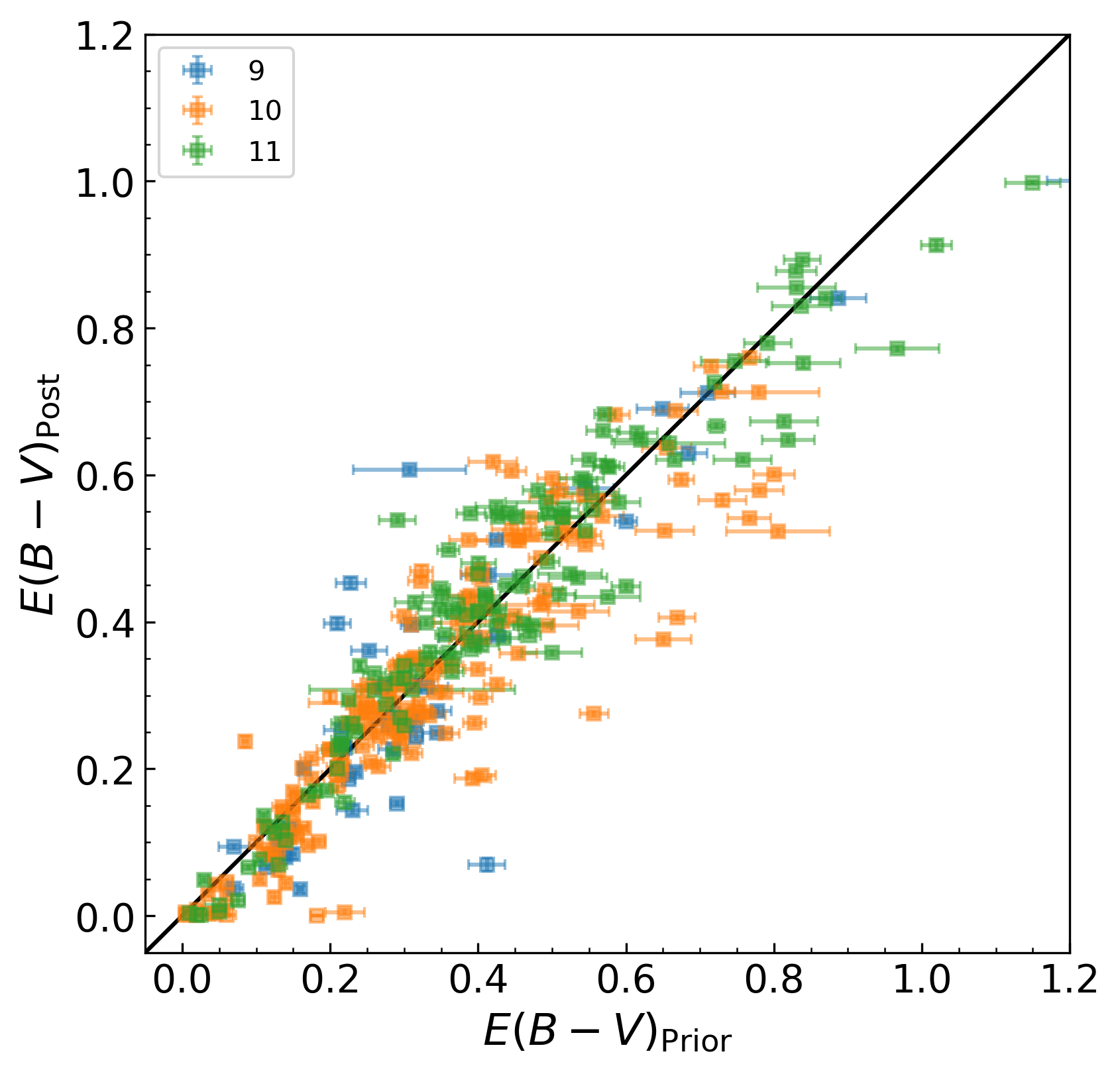}
    \caption{Estimation of $E(B-V)$ utilizing multiband $P$-$L$ relations of $\delta$ Scuti stars on held-out test sources. Blue, orange and green squares represent sources with 9, 10 and 11 photometric measurements, respectively. The x-axis shows the $E(B-V)$ values derived from the 3D dust map by \cite{green20193d}, while the y-axis displays the corresponding estimates obtained from the application of our model, using an uninformative color excess prior.}
    \label{fig:pred_ebv}
\end{figure}

\begin{figure}
    \centering
    \includegraphics[width=0.5\textwidth]{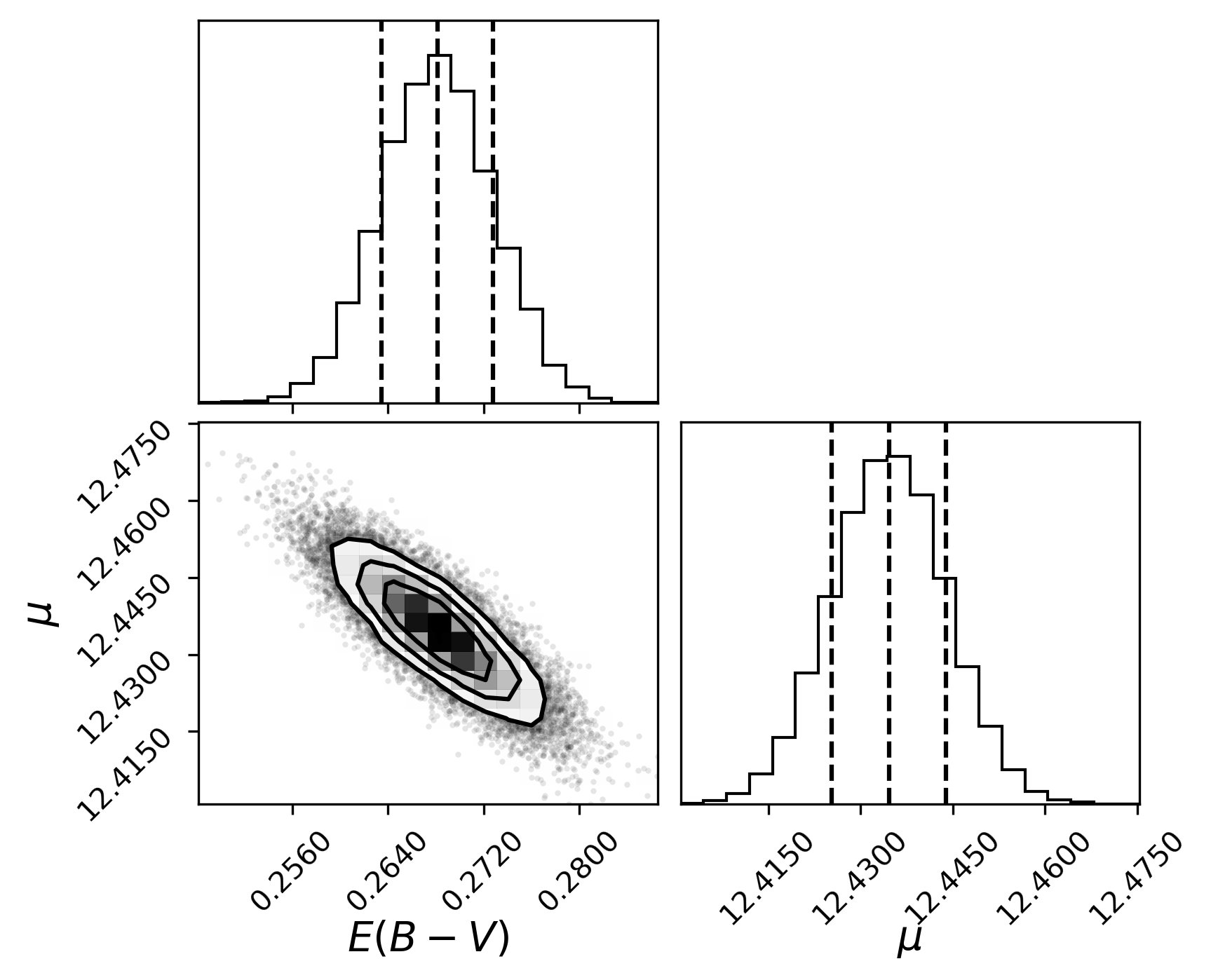}
    \caption{Contour plots for the joint posterior distributions of distance modulus and $E(B-V)$ for TMTS J00572464+5750191. The expected anti-correlation between extinction and distance is evident.}
    \label{fig:contour_test}
\end{figure}

\begin{figure}
    \centering
    \includegraphics[width=0.5\textwidth]{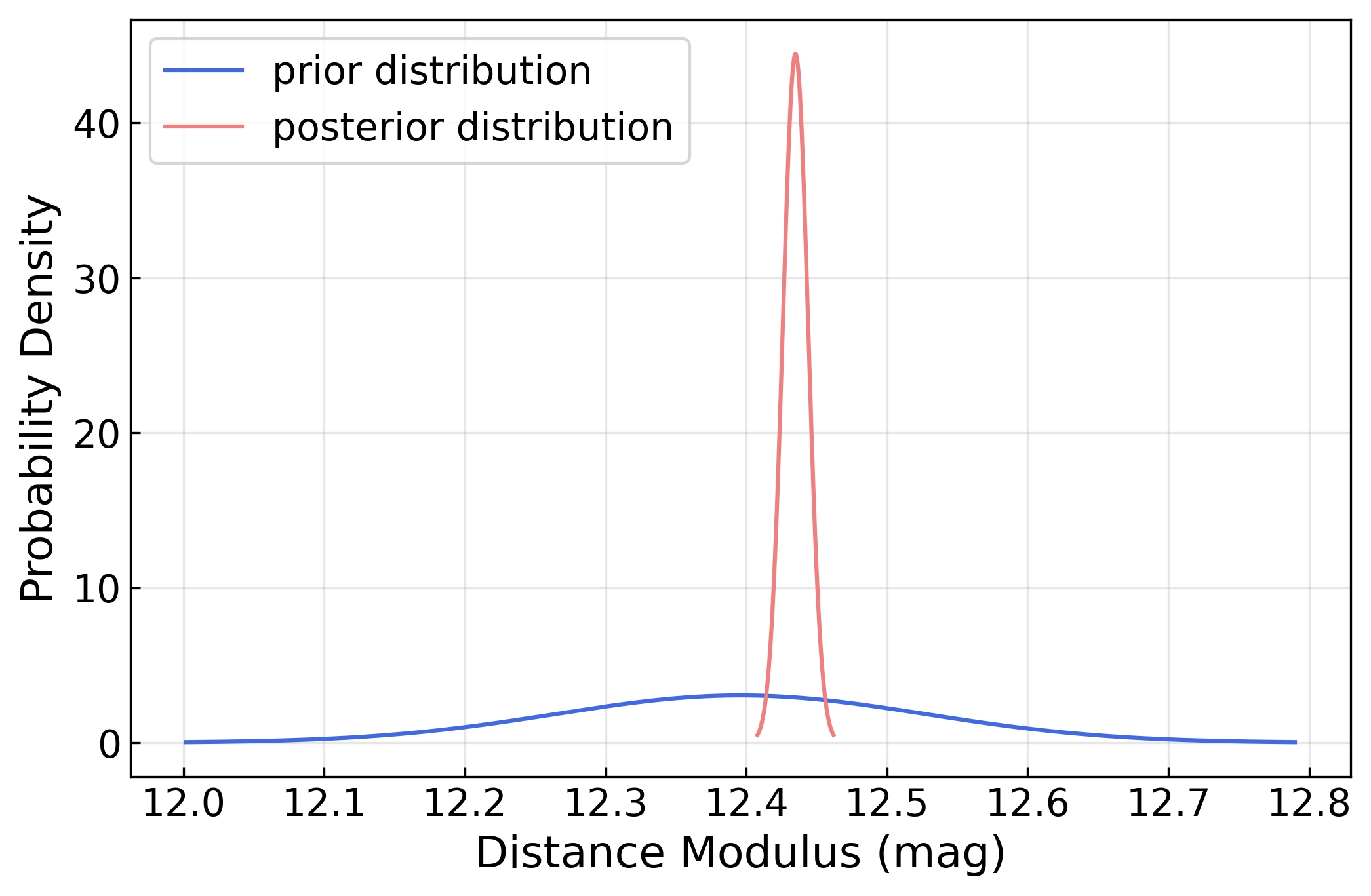}
    \caption{Comparison of the prior distance modulus distribution, derived from $Gaia$ DR3, and the posterior distribution, demonstrating a substantial improvement in precision in the latter.}
    \label{fig:distance_pos}
\end{figure}

\begin{figure*}
    \centering
    \includegraphics[width=1\textwidth]{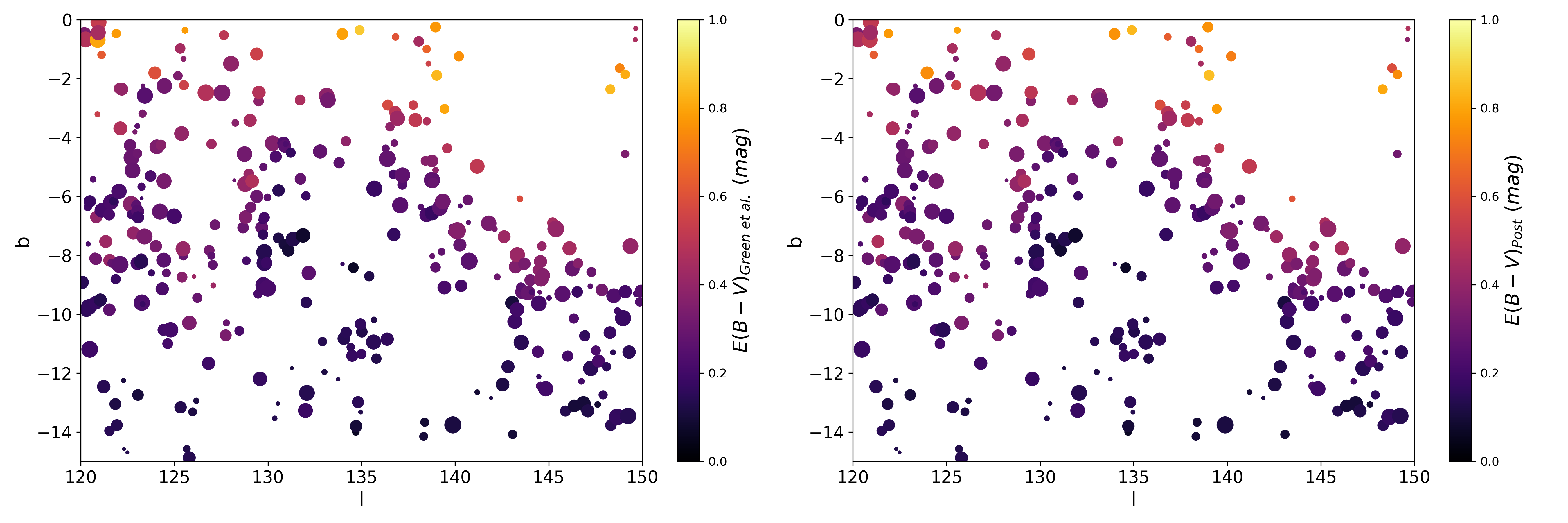}
    \caption{The $E(B-V)$ values between 1 and 3 kpc estimated from the dust map of \cite{green20193d} (left) and $\delta$ Scuti stars in our dataset (right). The size of each data point represents the distance to the source, with larger points corresponding to closer sources.}
    \label{fig:ebv}
\end{figure*}

To enhance the reliability of our results, we focus exclusively on variables with photometric measurements in at least nine bands. As shown in Figure \ref{fig:pie}, approximately 60\% of the sources in our dataset fulfill this criterion. Figure \ref{fig:pred_ebv} compares the posterior distribution of $E(B-V)$ with the values obtained from \texttt{DUSTMAPS} (version = "bayestar19", \citealt{green20193d}), showing a close alignment between the two. This shows that we are able to estimate $E(B-V)$ independently without clear bias, highlighting the reliability of our framework. 

Note that the $P$-$L$ relations of $\delta$ Scuti stars are not perfectly linear due to various reasons discussed above. As a result, the estimation of $E(B-V)$ based on these relations is subject to uncertainties arising from this imperfection. Consequently, the dust maps produced using our approach may not achieve the same level of precision as those given by \cite{green20193d}. However, the intrinsic scatter is relatively small in most bands, suggesting that, while various factors related to the physical properties of $\delta$ Scuti stars contribute to the dispersion of the $P$-$L$ relations, their impact remains within an acceptable range.

In addition, incorporating multiband photometry strengthens our constraints on dust extinction compared to single-band $P$-$L$ relation-based methods. Since dust extinction is the only free parameter for a given $\delta$ Scuti star during the fitting process, as in Equation \ref{eq:plr-3}, increasing the number of photometric bands enhances the observational constraints because of reducing degrees of freedom. Therefore, despite the existence of intrinsic scatter in the $P$-$L$ relations, our extinction estimates remain reliable within meaningful limits.

Despite these limitations, our method offers a unique advantage---simultaneous inference of both distance modulus and dust extinction, enabling us to extend the dust map beyond the limits of $Gaia$ parallax measurements. The 3D dust maps rely on distance measurements, with most current dust maps being primarily based on $Gaia$ parallaxes, which remain reliable ($\varpi/\sigma_\varpi \geq 10.0$) only within a few kpc. For sources lacking reliable $Gaia$ parallax measurements (e.g., $10.0 \geq \varpi/\sigma_\varpi \geq 5.0$, or even worse), we can constrain their distances and dust extinction through the multiband $P$-$L$ relations. As an example, we simultaneously estimated the distance modulus and dust extinction for TMTS J00572464+5750191, a source with $\varpi/\sigma_\varpi \approx 7$. Figure \ref{fig:contour_test} presents the contour plot for the joint posterior distributions of the distance modulus and $E(B-V)$ for this source, and Figure \ref{fig:distance_pos} compares the prior and posterior distribution of its distance modulus, demonstrating a significantly improved constraint. 

Since $\delta$ Scuti stars are widely distributed throughout the Galaxy, and the detection depth of large-scale surveys such as LSST (5$\sigma$ depth for $r$-band = 24.3 mag, \citealt{ivezic2019lsst}) and the Wide Field Survey Telescope (WFST, 5$\sigma$ depth for $r$-band > 21 mag, \citealt{lin2024minute}) has greatly increased, multiband photometry of distant $\delta$ Scuti stars is becoming accessible. Their $P$-$L$ relations would offer an independent means to refine measurements of distance modulus and dust extinction. This capability enables us to expand the spatial coverage of 3D dust maps to more distant regions, where $Gaia$ parallaxes become less reliable, highlighting the critical role of pulsating stars in constructing more comprehensive dust maps.

In the future, the large number of $\delta$ Scuti stars identified by LSST will facilitate the construction of 3D dust map through their $P$-$L$ relations. The single-visit depth of LSST is 24.3 mag, and the typical absolute magnitude is $M_{0,r} = 1.83$ mag. As such, the extinguished maximum distance at which LSST can detect $\delta$ Scuti stars is 312 kpc, which far exceeds the stellar locus Galaxy. Even with 7.5 mag of $r$-band extinction towards the Galactic plane, $\delta$ Scuti stars can be detected by LSST to $\sim$ 8 kpc (though stellar crowding may ultimately be limiting). Roughly scaling from the total number of stars expected to be detected in LSST, we estimate that LSST will ultimately discover and characterize a few $\times 10^6$ $\delta $ Scuti stars. Figure \ref{fig:ebv} compares the dust extinction derived from the map of \cite{green20193d} with $\delta$ Scuti stars in our dataset, focusing on the range between 1 and 3 kpc. 

Furthermore, it is worth noting that as this approach is inherently based on the $P$-$L$ relations of pulsating stars, it is not restricted to $\delta$ Scuti stars; but is broadly applicable to any pulsating stars that follow the $P$-$L$ relations, including Cepheids and RR Lyrae stars, whose $P$-$L$ relations are tighter and better-studied. Given the widespread presence of pulsating stars across the Galaxy, this underscores the potential of our method to refine and expand dust maps into the outer regions of the Galaxy. This capability offers great opportunity for studying the three-dimensional structure of interstellar dust in the Milky Way, offering insights into the spatial variation of extinction and the large-scale properties of the interstellar medium. 

Leveraging pulsating stars as distance and extinction tracers, our 3D dust maps enable the exploration of more distant regions, greatly broadening their applicability. For example, we might be able to explore low-surface-brightness substructures with these comprehensive dust maps, such as Milky Way satellite galaxies with tidal tails \citep{bournaud2004kinematics}, which arise from galaxy interactions. The mapping of dust and gas distributions offers critical insights into the dynamical and chemical characteristics of these substructures, shedding light on their formation mechanisms, the influence of tidal forces and the distribution of dark matter \citep{hozumi2015development}. Furthermore, dust maps covering greater distances facilitates the identification of obscured star-forming regions within these systems \citep{knierman2013tidal}, enabling a deeper investigation of the interactions among dust, gas, and dark matter that govern their evolutionary pathways.

% -------------------------------------------------------
\section{Period-Luminosity-Metallicity relations}
\label{sec:plmr}
% -------------------------------------------------------

\begin{figure}
    \centering
    \includegraphics[width=0.5\textwidth]{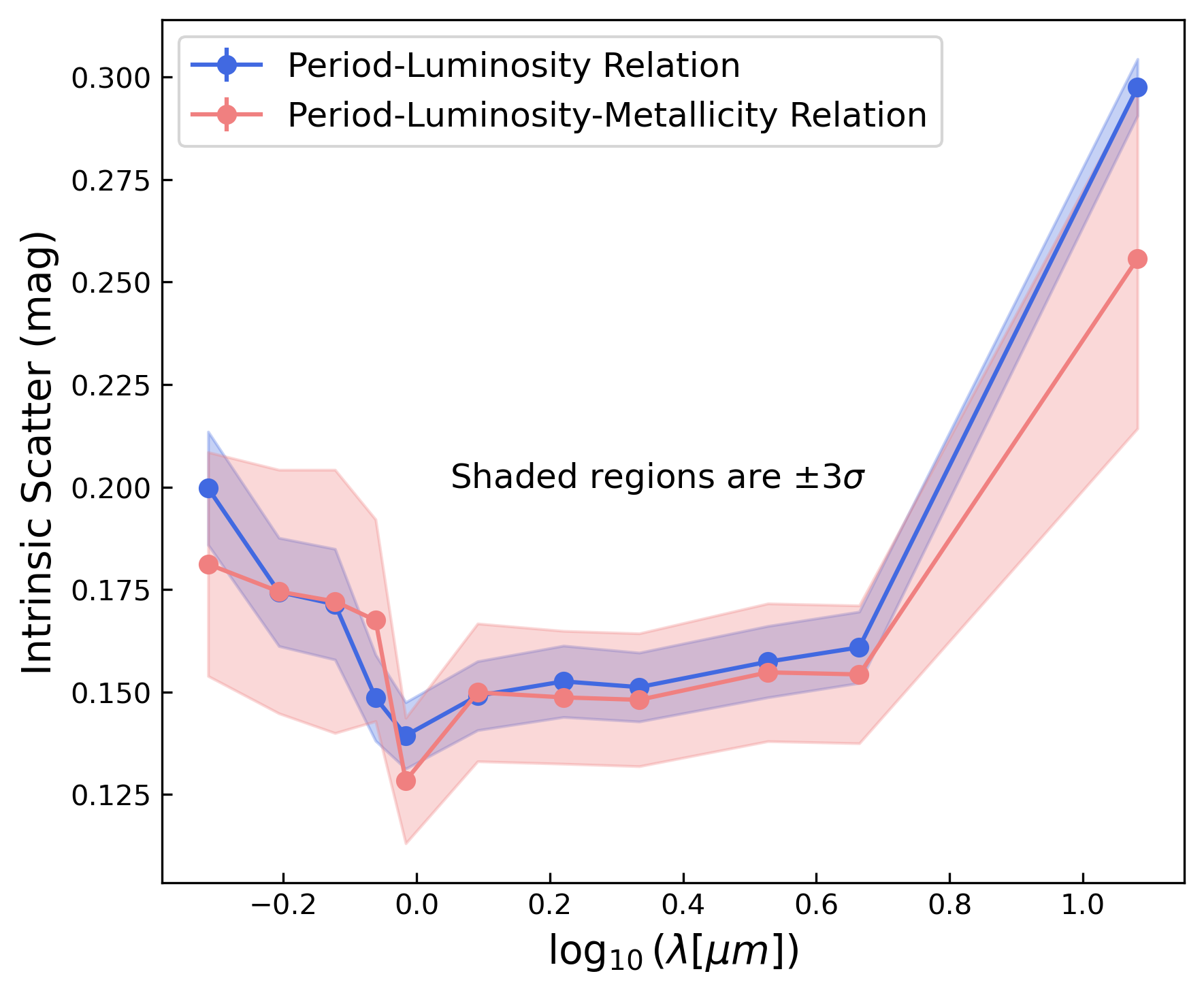}
    \caption{Intrinsic scatter in the $P$-$L$ relation (blue) and $P$-$L$-$Z$ relation (red) as a function of wavelength. Incorporating metallicity appears to reduce the intrinsic scatter at longer wavelengths.}
    \label{fig:scatter_com}
\end{figure}

The study of metallicity influence on the $P$-$L$ relations of pulsating stars is important for enhancing the precision of various astrophysical measurements, as metallicity represents a potential source of uncertainty in the $P$-$L$ relations \citep{de2022updated}. Metallicity dependence can affect distance determinations and the reconstruction of reliable dust maps across different regions of the Galaxy, highlighting the need to account for these effects.

Some studies have analyzed the effect of metallicity on the $P$-$L$ relations of $\delta$ Scuti stars (i.e., the $P$-$L$-$Z$ relations). For example, \cite{antonello1997luminosity} investigated the impact of metallicity on the $P$-$L$ relations of $\delta$ Scuti stars and concluded that these relations remain consistent regardless of metallicity, in contrast to the findings of \cite{nemec1995sx}. \cite{cohen2012sx} incorporated SX Phe stars (a subtype of $\delta$ Scuti stars with lower metal abundances) into the determination of the $P$-$L$ relations. Their results demonstrated great consistency with the relations derived from metal-rich pulsators, suggesting that metallicity has a minimal impact on the $P$-$L$ relations. \cite{mcnamara2011delta} treated the influence of metallicity as a correction term to absolute bolometric magnitudes, incorporating a linear metallicity-dependent term into the $P$-$L$ relations to account for its impact.

Theoretically, metals contribute to the opacity in the outer layer of a star, affecting the radiative transfer processes within the star, which can modify the  internal structure of the pulsator and nuclear fusion rates \citep{lenz2008asteroseismic,montalban2008input}. Metal-rich stars are generally less luminous and cooler compared to metal-poor stars at the same age and mass \citep{mowlavi1998some}, thus introducing a metallicity dependence in the $P$-$L$ relation.

We investigated the influence of metallicity on the $P$-$L$ relations using the metal abundance value obtained from LAMOST DR7. Among the $\delta$ Scuti stars in our dataset, 494 have [Fe/H] measurements from LAMOST. Following the approach of \cite{mcnamara2011delta}, we incorporated the effect of metallicity as a linear term into Equation \ref{eq:plr-2}, resulting in the $P$-$L$-$Z$ relation:

\begin{equation}
\begin{split}
    m_{i,j}=\mu_i + M_{0,j} + \alpha_j \log_{10}(P_i/P_0) + E(B-V)_i(a_jR_v+b_j) \\
    + \beta_j[Fe/H]_i +\epsilon_{i,j}.
	\label{eq:plmr}
\end{split}
\end{equation}
Where $\beta$ represents the coefficient of the metallicity term. As summarized in Table \ref{tab:plmr}, metallicity appears to exert a minimal influence on the $P$-$L$ relations of $\delta$ Scuti stars. All $\beta$ coefficients remain close to zero within 2$\sigma$ significance, and are smaller than the intrinsic scatter of their respective photometric bands, indicating that the impact of metallicity is statistically insignificant and difficult to separate from other contributing factors, such as photometric uncertainties and the inherent scatter in the $P$-$L$ relations. These findings suggest that not including metallicity in the determination of $P$-$L$ relations might not compromise precision, aligning with the results of many previous studies.

We conducted a comparison of the intrinsic scatter between the $P$-$L$ relations and the $P$-$L$-$Z$ relations across various photometric bands, as shown in Figure \ref{fig:scatter_com}. Our results indicate that incorporating metallicity seems to reduce the intrinsic scatter at longer wavelengths---as shown in Figure \ref{fig:scatter_com}, where the red line lies below the blue line in $H$, $K_s$, $W1$, $W2$, and $W3$ bands, typically by less than 0.01 mag. At shorter wavelengths, however, the impact of metallicity is less consistent, with the intrinsic scatter showing both increases and decreases. However, the observed effects of metallicity fail to achieve a 3$\sigma$ significance level in almost all bands (as illustrated by the shaded region of Figure \ref{fig:scatter_com}), suggesting that the trend may not be statistically robust. Furthermore, all the $\beta$ remain smaller than the intrinsic scatter itself, highlighting the possibility that the observed changes could be attributed to statistical fluctuations rather than to the physical impact of metallicity on the $P$-$L$ relations. More studies are required to understand the impact of metallicity on $\delta$ Scuti stars, including larger sample sizes and more detailed modeling of other confounding factors, such as the influence of age, binarity, or rotational effects on pulsation properties. 

\begin{table*}
	\centering
	\caption{$\alpha$, $M_0$, $\beta$ and intrinsic scatters of the 11-band $P$-$L$-$Z$ relations.}
	\label{tab:plmr}
	\begin{tabular}{ccccc} 
		\hline
		Band & $\alpha$ & $M_0$ & $\beta$ & $\sigma_\mathrm{{intrinsic}}$ (mag)\\
		\hline
		g & $-$2.8232$\pm$0.0934 & 1.9612$\pm$0.0122 & 0.0512$\pm$0.0303 & 0.181$\pm$0.009\\
		r & $-$3.0773$\pm$0.1013 & 1.8553$\pm$0.0134 & 0.0327$\pm$0.0247 & 0.175$\pm$0.010\\
            i & $-$3.0934$\pm$0.1064 & 1.9005$\pm$0.0144 & 0.0294$\pm$0.0232 & 0.172$\pm$0.011\\
            z & $-$3.0994$\pm$0.0862 & 1.9394$\pm$0.0115 & 0.0172$\pm$0.0147 & 0.168$\pm$0.008\\
            y & $-$3.2676$\pm$0.0513 & 1.8843$\pm$0.0069 & 0.0083$\pm$0.0076 & 0.128$\pm$0.005\\
            J & $-$3.3784$\pm$0.0567 & 1.1389$\pm$0.0074 & 0.0078$\pm$0.0073 & 0.150$\pm$0.006\\
            H & $-$3.4539$\pm$0.0561 & 1.0144$\pm$0.0074 & 0.0080$\pm$0.0074 & 0.149$\pm$0.005\\
            K$_s$ & $-$3.4536$\pm$0.0557 & 0.9825$\pm$0.0073 & 0.0074$\pm$0.0070 & 0.148$\pm$0.005\\
            W1 & $-$3.4613$\pm$0.0579 & 0.9625$\pm$0.0077 & 0.0081$\pm$0.0075 & 0.155$\pm$0.006\\
            W2 & $-$3.4681$\pm$0.0571 & 0.9873$\pm$0.0076 & 0.0087$\pm$0.0080 & 0.154$\pm$0.006\\
            W3 & $-$3.3245$\pm$0.1401 & 0.8989$\pm$0.0170 & 0.0754$\pm$0.0518 & 0.256$\pm$0.014\\
		\hline
	\end{tabular}
\end{table*}

% -------------------------------------------------------
\section{Summary}
\label{sec:summary}
% -------------------------------------------------------

With a cadence of 1 minute, the TMTS has proven to be a powerful tool for identifying short-period variable stars. Leveraging the dataset of 1,864 fundamental mode $\delta$ Scuti stars from the TMTS catalogs of Periodic Variable Stars, we developed an innovative methodology to simultaneously determine the multiband $P$-$L$ relations of these pulsators. By cross-matching TMTS sources with Pan-STARRS1, 2MASS, and WISE for optical and infrared photometry, our approach has achieved highly precise $P$-$L$ relations spanning multiple photometric bands. Additionally, this methodology facilitates the refinement of the distance modulus and color excess within the posterior distribution.

A notable application of this methodology lies in its capability to estimate the color excess independently. Derived without prior information of dust extinction, our $E(B-V)$ estimates are in good agreement with results from state-of-the-art 3D dust maps. This demonstrates the utility of our approach as a complementary method for extinction estimation. This extensible method also offers a potential to estimate dust extinction with other types of pulsating stars that follow well-studied $P$-$L$ relations, such as Cepheids and RR Lyrae stars. When parallax measurements are not reliable or even unavailable, this method can also be used to determine distance to $\delta$ Scuti stars. More numerous than Cepheids and RR Lyrae, $\delta$ Scuti stars could become the de facto sources for distance measurements to Galactic substructures, like tidal tails.

We also investigated the influence of metallicity on the $P$-$L$ relations of $\delta$ Scuti stars by incorporating it as a linear term in the fitting. Our findings suggest that the inclusion of metallicity may reduce intrinsic scatter at longer wavelengths, potentially refining the precision of the $P$-$L$ relations in these bands. However, the observed effects do not reach 3$\sigma$ statistical significance, with the metallicity coefficients remaining within the intrinsic scatter of the respective bands, making it challenging to distinguish these effects from statistical fluctuations. Further investigation is required to conclusively evaluate the physical impact of metallicity on these relations.

% -------------------------------------------------------
\begin{acknowledgement}
This work is supported by the National Science Foundation of China (NSFC grants 12033003 and 12288102), the Ma Huateng Foundation and New Cornerstone Science Foundation through the XPLORER PRIZE. J.L. is supported by the National Natural Science Foundation of China (NSFC; Grant Numbers 12403038), the Fundamental Research Funds for the Central Universities (Grant Numbers WK2030000089), and the Cyrus Chun Ying Tang Foundations.\\

This work is supported by the National Natural Science Foundation of China (NSFC) grant 12373031, the Joint Research Fund in Astronomy (U2031203) under cooperative agreement between the National Natural Science Foundation of China (NSFC) and Chinese Academy of Sciences (CAS), and the NSFC grants 12090040, 12090042. This work is also supported by the the CSST project (CCST-A12).\\

We acknowledge the support of the staffs from Xinglong Observatory of NAOC during the installation, commissioning, and operation of the TMTS system. \\

This work has made use of data from the European Space Agency (ESA) mission
{\it Gaia} (\url{https://www.cosmos.esa.int/gaia}), processed by the {\it Gaia}
Data Processing and Analysis Consortium (DPAC,
\url{https://www.cosmos.esa.int/web/gaia/dpac/consortium}). Funding for the DPAC
has been provided by national institutions, in particular the institutions
participating in the {\it Gaia} Multilateral Agreement.\\

The Pan-STARRS1 Surveys (PS1) and the PS1 public science archive have been made possible through contributions by the Institute for Astronomy, the University of Hawaii, the Pan-STARRS Project Office, the Max-Planck Society and its participating institutes, the Max Planck Institute for Astronomy, Heidelberg and the Max Planck Institute for Extraterrestrial Physics, Garching, The Johns Hopkins University, Durham University, the University of Edinburgh, the Queen's University Belfast, the Harvard-Smithsonian Center for Astrophysics, the Las Cumbres Observatory Global Telescope Network Incorporated, the National Central University of Taiwan, the Space Telescope Science Institute, the National Aeronautics and Space Administration under Grant No. NNX08AR22G issued through the Planetary Science Division of the NASA Science Mission Directorate, the National Science Foundation Grant No. AST-1238877, the University of Maryland, Eotvos Lorand University (ELTE), the Los Alamos National Laboratory, and the Gordon and Betty Moore Foundation.\\

This publication makes use of data products from the Two Micron All Sky Survey, which is a joint project of the University of Massachusetts and the Infrared Processing and Analysis Center/California Institute of Technology, funded by the National Aeronautics and Space Administration and the National Science Foundation.\\

This publication makes use of data products from the Wide-field Infrared Survey Explorer, which is a joint project of the University of California, Los Angeles, and the Jet Propulsion Laboratory/California Institute of Technology, funded by the National Aeronautics and Space Administration.\\

Guoshoujing Telescope (the Large Sky Area Multi-Object Fiber Spectroscopic Telescope LAMOST) is a National Major Scientific Project built by the Chinese Academy of Sciences. Funding for the project has been provided by the National Development and Reform Commission. LAMOST is operated and managed by the National Astronomical Observatories of the Chinese Academy of Sciences.
\end{acknowledgement}

\bibliography{references}

\end{document}